\documentclass[11pt]{article}
  
  \newcommand{\nc}{\newcommand}
   \nc{\wh}{\widehat}
   \nc{\pl}{\partial}
   
   \nc{\inv}{^{-1}}
  \def\ra{\rightarrow}
  \def\iy{\infty}

  \def\be{\begin{equation}}
  \def\ee{\end{equation}}
  \def\ba{\begin{eqnarray*}}
  \def\ea{\end{eqnarray*}}
  \def\bae{\begin{eqnarray}}
  \def\eae{\end{eqnarray}}
  \def\bc{\begin{center}}
  \def\ec{\end{center}}
  \def\ov{\over}
  \def\al{\alpha}

  \def\la{\lambda}
  \def\pr{\textrm{Prob}}
  
  \def\cd{\cdots}
  \def\Dl{\Delta}
  \def\dl{\delta}
  \def\cA{\mathcal{A}}
  
  \def\x{\xi}
  
\begin{document}
  \title{Random Words, Toeplitz Determinants and
  Integrable Systems. II. }
  \date{October 20, 2000}
  \author{
  Alexander R.~Its\\
  Department of Mathematics\\
  Indiana University-Purdue University Indianapolis\\
  Indianapolis, IN 46202, USA
  \and
  Craig A.~Tracy\\
  Department of Mathematics\\
  Institute of Theoretical Dynamics\\
  University of California\\
  Davis, CA 95616, USA
  \and
  Harold Widom\\
  Department of Mathematics\\
  University of California\\
  Santa Cruz, CA 95064, USA
  }
  \maketitle
  \section{Introduction}
  This paper, a continuation of \cite{itw1},   connects   the analysis of
  the length of the
  longest weakly increasing subsequence of inhomogeneous random
  words to   a Riemann-Hilbert problem and an associated system of integrable
  PDEs.    That such a connection exists is not so surprising given the
fundamental
  work of Baik, Deift and Johansson~\cite{bdj1} connecting the related problem
  involving random permutations to a Riemann-Hilbert problem.
  For the reader's convenience
  we first summarize some of the results of
  \cite{itw1} before presenting our new results.

  A word is  a string of  symbols, called \textit{letters}, which
  belong to an ordered alphabet $\cA$ of fixed size $k$.  The set of
  all such words  of length $N$,  $\mathcal{W}(\cA,N)$, forms the sample space
in
   our statistical
  analysis. We equip the space $\mathcal{W}(\cA,N)$ with a natural {\it
inhomogeneous\/}
  measure  by assigning to each letter $i \in \cA $ a probability $p_{i}$ and
defining
  the probability measure on  words by the product measure. We also order the
$p_i$ so that
  \[ p_1\ge p_2 \ge \cdots \ge p_k \]
  and decompose our alphabet $\cA$ into subsets $\cA_1$, $\cA_2$, \ldots,
  $\cA_M$, $M\leq k$, such
  that $p_i=p_j$ if and only if $i$ and $j$ belong to the same $\cA_\al$.

  Let
  \[
  w=\al_1\al_2\cdots \al_N\in \mathcal{W}, \quad \al_{i} \in \cA,
  \]
  be a word. A {\it weakly increasing subsequence\/} of the word $w$
    is a subsequence $\al_{i_1}\al_{i_2}\cdots \al_{i_m}$
  such that $i_1<i_2<\cdots<i_m$ and $\al_{i_1}\le \al_{i_2}\le \cdots \le
\al_{i_m}$.
  The positive integer $m$ is called the \textit{length} of this weakly
increasing
  subsequence.
  For each word $w\in\mathcal{W}$ we define $\ell_N(w)$ to equal the
\textit{length of
  the longest
  weakly increasing subsequence\/} in $w$.\footnote{There may be many
subsequences of $w$
  that have the
  identical length $\ell_N(w)$.}
  The function
  \[
  \ell_N : \mathcal{W}(\cA,N)\mapsto \textbf{R}
  \]
   is the principal random variable in our analysis, and
  the corresponding distribution function,
   \[ F_N(n):=\textrm{Prob}\left(\ell_N(w)\le n\right), \]
   is our principal object.
  ($\textrm{Prob}$ is the   inhomogeneous measure on random words; it
  depends  upon $N$ and the probabilities $p_i$.)

  To formulate the basic result of \cite{itw1},   define
  \[
  k_{\al} = |\cA_{\al}|,
  \]
  where
  \[
  \cA = \bigcup_{\al = 1}^{M}\cA_{\al}
  \]
  is the decomposition of the alphabet $\cA$  introduced above, then

  \ba \lim_{N\ra\iy}\pr\left({\ell_N - N p_1 \ov \sqrt{N p_1}}\le s\right)
  &=&(2\pi)^{-(k-1)/2}\prod_\al (1!\,2!\cd (k_\al-1)!)^{-1}\, \times \\
  & &\raisebox{-3mm}{$\displaystyle
  \int\cdots\int \atop \hspace{-2mm}
  \textstyle{{\xi_i \in\, \raisebox{-.3ex}{$\displaystyle{\Xi}$} \atop
  \vspace{.5ex} \xi_1\le s}}$} \prod_\al
  \Dl_\al(\x)^2\;
  e^{-\sum\x_i^2/2}\;\dl(\sum\sqrt{p_i}\,\x_i)\, d\xi_1\cdots d\xi_k,\ea
  where $\Dl_\al(\x)$ is the Vandermonde determinant of those $\x_i$
  with $i \in \cA_{\al}$, and $\Xi$ denotes the set of those $\x_i$
  that $\x_{i+1} \leq \x_{i}$ whenever $i$ and $i+1$ belong to the
  same $\cA_{\al}$.

  This result has the following random matrix interpretation.
  The limiting distribution function (as $N\ra \iy$)  for
  the appropriately centered and normalized random variable $\ell_N$ is related
to the
  distribution function for the eigenvalues $\xi_i$ in the \textit{direct sum}
of
   mutually independent $k_\al\times k_\al$ Gaussian unitary
  ensembles,\footnote{A basic reference for random matrices is Mehta's
   book~\cite{mehta}.}
  conditional on the eigenvalues $\xi_i$ satisfying $\sum\sqrt{p_i}\,\xi_i=0$.
  In the case when one letter occurs with greater probability than the others,
this
  result implies
  that  the limiting distribution of $(\ell_N-N p_1)/\sqrt{N}$ is Gaussian with
   variance equal to $p_1(1-p_1)$.
  In the  case when all the probabilities are distinct,
  we proved the refined asymptotic result
  \[\textrm{E}(\ell_N)=Np_1+\sum_{j>1}{p_j\ov p_1-p_j}+O({1\ov \sqrt N}),
  \> N\ra\iy.\]

  The derivation of the above asymptotic formulae follows from a
  direct asymptotic analysis of the right hand side of the  basic combinatorial
equation,
  \[ \pr\left(\ell_N(w)\le n\right)=\sum_{{\la\vdash N \atop \la_1\le n}}
  s_\la(p) \, f^\la.\]
  Here $\la \vdash N$ denotes a
  partition of $N$,
   $s_{\la}(p)$ is the Schur function of shape $\la$
  evaluated at $p:= (p_1, p_2, ..., p_k, 0, 0,...)$, and $f^\la$
  equals the number of standard Young tableaux  of shape $\la$, see, e.g.\
  \cite{stanleyBook}.
  After \cite{itw1} was written, Stanley~\cite{stanley}  showed that
  the measure $\pr(\{\la\}):=s_\la(p) f^\la$ also underlies the analysis of
certain
  (generalized) riffle shuffles of Bayer and Diaconis~\cite{bayer}.
  Stanley relates this measure to quasisymmetric functions and does
  not require that $p$ have finite support.  (Many of our results
  generalize to the case when $p$ does not have finite support, but we
  do not consider this here.)  The measure considered here and in~\cite{stanley}
  is  a specialization of the Schur measure 
  $\pr(\{\la\}):=s_\la(x) s_\la(y)$~\cite{okounkov}.  
  For the Schur measure, Okounkov~\cite{okounkov}
  has shown that the associated correlation functions satisfy an infinite
  hierarchy of PDEs; namely, the Toda lattice hierarchy of 
  Ueno and Takasaki~\cite{ueno}.  Similar results were also obtained
  by Adler and van Moerbeke~\cite{adler1, vanM}.

  Gessel's theorem \cite{gessel} (see also \cite{itw1, tw2}) implies that the
  (exponential) generating
  function of $\pr(\ell_N\le n)$ is a Toeplitz determinant\footnote{If $\phi$
  is a function on the unit circle with Fourier coefficients $\phi_j
  := 1/2\pi \int_{-\pi}^{\pi}e^{-ij\theta}\phi(e^{i\theta})d\theta$ then
  $T_n(\phi)$ denotes the Toeplitz matrix
$\left(\phi_{i-j}\right)_{i,j=0,1,\ldots,n-1}$
  and $D_n(\phi)$ its determinant.}
  \be
   G_I(n;\{p_i\},t):= \sum_{N=0}^\infty \pr\left(\ell_N(w)\le n\right) \,
  {t^N\ov N!}= D_n(f_I), \label{toeplitzDetII}\ee
  where
  \[ f_I(z) = e^{t/z}\, \prod_{j=1}^k (1+p_j z). \]
  Probabilistically,
  $G_{I}(n;\{p_i\}, t)$ is   the {\it Poissonization\/} of $\ell_N$. Similar
  Poissonizations have proved crucial in the analysis of
  the length of the longest increasing
  subsequences in  random permutations~\cite{aldous, johansson0, bdj1}
  (see also \cite{johansson1, johansson2, tw2}
  and references therein).

  In the present  paper we use  (\ref{toeplitzDetII}) to express
$G_{I}(n;\{p_i\}, t)$
  in terms of the  solution of a certain integrable system of nonlinear PDEs.
  Indeed, we show that
  $G_{I}(n;\{p_i\}, t)$ can be identified as the \textit{Jimbo-Miwa-Ueno
  $\,\tau$-function}~\cite{jmu, jimbo} corresponding to the (generalized)
  Schlesinger isomonodromy deformation equations
  of the $2\times 2$ matrix linear ODE which has $M+1$ simple poles in the
  finite complex plane and one
  Poincar\'e index 1 irregular singular point at infinity.  Recall that the
number $M$ is
  the total number of
  the  subsets  $\cA_\al \subset \cA$.  The poles are located at $0$ and
    $-p_{i_{\al}}$ ($i_{\al}$ = max $\cA_\al$). The integers
  $k_{\al}$ appear as the formal monodromy exponents at the respective
  points  $-p_{i_{\al}}$. We also evaluate the remaining  monodromy data and
  formulate a $2 \times 2$ matrix Riemann-Hilbert problem which provides yet
another
  analytic representation for the function $G_{I}(n;\{p_i\}, t)$. Similar
  to the problems considered in \cite{bdj1} and \cite{bi}, the Riemann-Hilbert
  representation of    $G_{I}(n;\{p_i\}, t)$ can be used for the
  further asymptotic analysis of the random variable $\ell_N(w)$ via the
Deift-Zhou
  method~\cite{dz}.
  In the {\it homogeneous\/} case, i.e.\ when $M = 1$, the system of
Schlesinger equations
  we obtain reduces to a special  case of Painlev\'e V equation. This  result
  was obtained earlier in \cite{tw2}.
  The exact formulation of the results indicated above is presented in Theorem
1 in \S 4.

  Our derivation of the differential equations for the function
$G_{I}(n;\{p_i\}, t)$
  follows a scheme well known in  soliton
  theory (see e.g.\ \cite{nmpz}) called the  \textit{Zakharov-Shabat dressing
method}. We
  are able to apply this scheme since there exists
  a  matrix Riemann-Hilbert problem associated to
  any Toeplitz determinant as was shown by Deift~\cite{deift}.
  For the reader's convenience this is derived
  in \S 2.

  The basic idea of the
  Riemann-Hilbert approach to Toeplitz
  determinants suggested in \cite{deift}
    is a representation of a Toeplitz determinant $D_{n}(\phi)$
  as a Fredholm determinant of an integral operator acting on $L_{2}(C)$,
$C$=unit circle,
   and belonging
  to a special {\it integrable\/} class which admits a Riemann-Hilbert
representation
  \cite{iiks}. Borodin and Okounkov~\cite{bo} (see also \cite{bw} for a
  simplified derivation  and \cite{johansson2}, \cite{boo} for a particular
case of $\phi$)
  found a different  Fredholm determinant representation for  $D_{n}(\phi)$.
  The Fredholm operator in this representation acts on $l_{2}(\{n, n+1, ...\})$
which makes
   the
  representation quite suitable for the analysis of the large $n$ asymptotics
of $D_{n}(\phi)$
  (see \cite{johansson1} \cite{johansson2}, \cite{boo}).
  Borodin \cite{borodin} subsequently observed
  that the discrete Fredholm representation of \cite{bo} involves a discrete
analog
  of the integrable kernels and can be supplemented by a discrete analog of
  the Riemann-Hilbert problem. (This is similar to the \textit{pure} soliton
  constructions in the theory of integrable PDEs~\cite{nmpz}.)

  We conclude this introduction by noting that   our derivation of integrable
  PDEs for the Toeplitz determinant $D_{n}(f_I)$  can be
  applied to any Toeplitz determinant whose symbol $\phi$ satisfies
  the condition,
  \[
  \frac{d}{dz}\log \phi(z) = \mbox{rational function of $z$}.
  \]
  This is one place where the finite support of $p$ is crucial.  It is an
interesting
  open problem, particularly in light of \cite{stanley}, to remove this
restriction.

  \section{Fredholm Determinant Representation of the Toeplitz Determinant
  and the Riemann-Hilbert Problem}
  \setcounter{equation}{0}

   Let $\phi(z)$ be a continuous function on the unit circle $C = \{|z|=1\}$
oriented
   in the counterclockwise direction.
  Let $n \in \textbf{N}$ and denote by
    $K_{n}(\phi)$ the integral operator acting  on $L_{2}(C)$ with kernel
  \be \label{inop}
  K_{n}(z, z') = {{z^{n}(z')^{-n} -1}\ov {z-z'}}\, \, {{1-\phi(z')}\ov 2\pi
i}\, .
  \ee
  It was shown in \cite{deift} that
  \be \label{tfred}
  D_{n}(\phi)  = \det (1-K_{n}(\phi)),
  \ee
  where the determinant on the right is a Fredholm determinant taken in
$L_{2}(C)$.  (Note that $K_n(z,z')$ has no singularity at $z=z'$.)
  Equation (\ref{tfred}) follows from the ``geometric sum form''
of the kernel $K_n$,
\[ K_n(z,z')=\sum_{k=0}^{n-1} z^k\,{1-\phi(z')\ov 2\pi i}\, (z')^{-k-1} ,\]
 which shows that  
 the Toeplitz matrix $T_{n}(\phi)$ is essentially the matrix
representation
  of the operator $1-K_{n}(\phi)$ in the basis $\{z^k\}_{-\infty <k<\infty}$.
(For more details see \cite{deift}.)

  The integral operator $K_{n}(\phi)$ belongs to the class of
\textit{integrable
  Fredholm operators} \cite{iiks, tw3, deift}, i.e.,  its
  kernel is of the form
  \[
  K_{n}(z, z') = {f^{T}(z)g(z')\ov z-z'}\, ,
  \]
  where
  \be \label{f}
  f(z) = (f_{1}, f_{2})^{T} = (z^{n}, 1)^{T}
  \ee
  and
  \be \label{g}
  g(z) = (g_{1}, g_{2})^{T} = (z^{-n}, -1)^{T}\, {{1-\phi(z)}\ov 2\pi i}\, .
  \ee
  We require, so that there is no singularity on the diagonal of the kernel,
  \be f^T(z)\, g(z)=0. \label{nonsing}\ee
  An important property of  these operators  is
  that  the resolvent  $R_{n} =(1-K)^{-1}-1$ also belongs to the same class
  (see again  \cite{iiks}, \cite{tw3}, \cite{deift} ). Precisely,
  \be \label{resolvent}
  R_{n}(z, z') = {F^{T}(z)G(z')\ov z-z'},
  \ee
  where
  \[
  F_{j} = (1-K_{n})^{-1}f_{j}, \quad G_{j} = (1-K_{n}^{T})^{-1}g_{j}, \qquad
j=1,2\,.
  \]

  The vector functions $F$ and $G$  can be  in turn computed in terms
  of a certain matrix Riemann-Hilbert problem~\cite {iiks}.
  Indeed, let us define (cf.\ \cite{iiks, deift})  the $2\times 2$ matrix
valued function
  \be \label{Y}
  Y(z) = I - \int_{C}F(z')g^{T}(z'){dz'\ov z'-z}\quad z\notin C .
  \ee
  Let $Y_{\pm}(z)$ denote the boundary values of the function $Y(z)$ on the
  contour $C$,
  \[
   Y_{\pm}(z) = \lim_{{ z'\rightarrow z \atop z'\in \pm\textrm{{\scriptsize side}}}}Y_{+}(z').
  \]
  From (\ref{Y}) it follows that
  \be \label{Cauchyjump}
  Y_{+}(z) - Y_{-}(z) = -2\pi i F(z)g^{T}(z)
  \ee
  and hence  (recall (\ref{nonsing}))
  \[
  Y_{+}(z)f(z) =Y_{-}(z)f(z).
  \]
  Using this, the matrix identity
  \[
  F(z')g^{T}(z')f(z) = f^{T}(z)g(z')F(z'), \quad (\mbox{associativity of the
matrix
  product})
  \]
  and  (\ref{Y}), we have
  \[
  Y_{\pm}(z)f(z) = f(z)  - \int_{C} f^{T}(z)g(z')F(z'){dz'\ov z'-z} =
  f(z)  + \int_{C} K(z,z')F(z')dz'.
  \]
  From the definition of $F$
  it follows that
  \be \label{F}
  F(z) = Y_{\pm}(z)f(z).
  \ee
  This  and  (\ref{Cauchyjump})
  imply the jump equation,

  \be \label{jump}
  Y_{-}(z) = Y_{+}(z)(I + 2\pi i f^{T}(z)g(z)), \quad z \in C.
  \ee
  This equation, supplemented by the obvious analytic properties
  of the Cauchy integral in (\ref{Y}), shows
  that the function $Y(z)$ solves the following $2\times 2$ matrix
  Riemann-Hilbert problem:
  \begin{itemize}
  \item $Y(z)$ is holomorphic for all $z\notin C$,
  \item $Y(\infty) = I$,
  \item $ Y_{-}(z) = Y_{+}(z)H(z), \quad z \in C$,
  \end{itemize}
  where the jump matrix $H$ is
  \bae
  H(z) &=& I + 2\pi i f^{T}(z)g(z)\label{Hgen}\\
  &=& \left( \begin{array}{cc}
  2-\phi(z) & (\phi(z) -1)z^{n}\\
  (1-\phi(z))z^{-n} & \phi(z)
  \end{array}\right).\label{H}
  \eae
  These analytic properties  determine  $Y$ uniquely.
  To see this, we first observe that  $\det H(z) \equiv 1$ implies that the
  scalar function $\det Y(z)$ has no jump on $C$; and hence, it is holomorphic
  and bounded on the whole complex plane. This together with the
  normalization
  condition at $z = \infty$ implies that $\det Y(z) \equiv 1$. Suppose
  that $\tilde{Y}(z)$ is another solution. Since both the functions
  $\tilde{Y}(z)$ and  $Y(z)$ satisfy the same jump condition across
  the contour $C$, the matrix ratio $\tilde{Y}(z)Y^{-1}(z)$ has no jump
  across $C$. This means that  $\tilde{Y}(z)Y^{-1}(z) \equiv \textrm{constant}$, and
  from the condition at  $z = \infty$ we actually have that
  $\tilde{Y}(z)Y^{-1}(z)\equiv I$. The uniqueness now follows.

  Since $Y$ is the unique solution of the Riemann-Hilbert
  problem, one
  can  now reconstruct the resolvent $R$ using (\ref{F}) and the similarly
  derived identity
  \be \label{G}
  G(z) = (Y^{T}_{\pm})^{-1}(z)g(z),
  \ee
  for $G$.
  We shall refer to this Riemann-Hilbert
  problem  as the $Y$-RH problem.

  Following \cite{deift} the $Y$-RH problem can be transformed to an equivalent
  Riemann-Hilbert problem which is directly connected with the polynomials on
  the circle $C$ orthogonal with respect to the (generally complex) weight
  $\phi(e^{i\theta})$.
  To this end we first note that since the entries of $f$ are  polynomials in
$z$,
   (\ref{F}) implies that  $F$ is an entire function of $z$. Since $Y(z)
  \rightarrow I$ as $z\rightarrow \infty$, it follows in fact that
   $F$ is polynomial,
  \be \label{Fpol}
  F(z) = \left( \begin{array}{c}
             P_{n}(z)\\
             Q_{n-1}(z)
  \end{array} \right), \quad P_{n}(z) = z^{n} + ..., \quad Q_{n-1}(z) =
q_{n-1}z^{n-1} + ...,
  \ee
  for some constant $q_{n-1}$. On the other hand, denoting by $Y_{j}$ the
$j$-th column
  of the matrix $Y$, we obtain from the jump equation (\ref{jump})
  (or, more precisely, from the equation $Y_{+} = Y_{-}H^{-1}$)  that
  \bae
  Y_{1+}(z) &=& Y_{-}(z) \left( \begin{array}{c}
             \phi(z)\\
             (\phi(z) -1)z^{-n}
  \end{array} \right)\nonumber\\ [0.5ex]
  &=& -z^{-n} Y_{-}(z) \left( \begin{array}{c}
             0\\
             1
  \end{array} \right) +
  \phi(z)z^{-n}Y_{-}(z) \left( \begin{array}{c}
             z^{n}\\
             1
  \end{array} \right)\nonumber\\
  &=& -z^{-n} Y_{2-}(z) + \phi(z)z^{-n}Y_{-}(z)f(z)\nonumber \\
  &=& -z^{-n} Y_{2-}(z) + \phi(z)z^{-n}F(z). \label{jumpcolumn}
  \eae

  Define
  \[
  J(z) = \left\{\begin{array}{ll}
          -Y_{1}(z),&\quad |z|<1,\\
          z^{-n}Y_{2}(z),&\quad |z|>1,
         \end{array}\right.
  \]
  and consider the $2\times 2$ matrix function
  \be \label{Z}
  Z(z) = \sigma_{3}\left(F(z), J(z)\right)\sigma_{3},\quad
  \sigma_{3} = \left( \begin{array}{cc}
                1&0\\
                0&-1
              \end{array}\right).
  \ee
  The function $Z$ is analytic outside of $C$, and
  it has the following asymptotic behavior as $z\rightarrow \infty$:
  \be \label{Zas}
  Z(z) = \left(I + O\left({1\ov z}\right)\right)
         \left( \begin{array}{cc}
                z^{n}&0\\
                0& z^{-n}
              \end{array}\right).
  \ee
  For the jump relation on the contour $C$ we have from (\ref{jumpcolumn}),
  \[
  Z_{+}(z) = \sigma_{3}\left(F(z), -Y_{1+}(z)\right)\sigma_{3}
  = \sigma_{3}\left(F(z), z^{-n} Y_{2-}(z) - \phi(z)z^{-n}F(z)\right)\sigma_{3}
  \]

  \[
  = \sigma_{3}\left(F(z), z^{-n}Y_{2-}(z)\right)
  \left( \begin{array}{cc}
                1 & - \phi(z)z^{-n}\\
                0& 1
              \end{array}\right)\sigma_{3}
  = \sigma_{3}\left(F(z), z^{-n}Y_{2-}(z)\right)\sigma_{3}
  \sigma_{3}\left( \begin{array}{cc}
                1 & - \phi(z)z^{-n}\\
                0& 1
              \end{array}\right)\sigma_{3}
  \]

  \[
  = Z_{-}(z)
  \left( \begin{array}{cc}
                1 & \phi(z)z^{-n}\\
                0& 1
              \end{array}\right).
  \]
  Summarizing the analytic properties of  $Z$ we
  conclude that it solves the following Riemann-Hilbert problem:
  \begin{itemize}
  \item $Z$ is holomorphic for all $z\notin C$,
  \item $Z(z)z^{-n\sigma_{3}} \rightarrow I,\quad z\rightarrow \infty$,
  \item $ Z_{+}(z) = Z_{-}(z)S(z), \quad z \in C$,
  \end{itemize}
  where the jump matrix  $S(z)$ is
  \be \label{Sz}
  S(z) = \left( \begin{array}{cc}
  1 & z^{-n}\phi(z)\\
  0 & 1
  \end{array}\right).
  \ee
  We shall refer to this Riemann-Hilbert problem as $Z$-RH problem.
  As in the $Y$-RH problem, the solution of the $Z$-RH problem
  is unique. Indeed, assuming that $\tilde{Z}$ is
  another solution, we introduce the matrix ratio $ X :=
  \tilde{Z}Z^{-1}$.
  By the same reasoning as in the case of the $Y$-RH problem, we
  conclude that $X$ is entire. Since
  \[
  X(z) = (\tilde{Z}(z)z^{-n\sigma_{3}})(z^{n\sigma_{3}} Z^{-1}(z))
  \rightarrow I\>\>\textrm{as}\>\> z\ra\iy,
  \]
  it follows that $X \equiv I$; and hence, that $Z$ is unique.
  We note that $Y$ (and hence the resolvent $R$) can be
  reconstructed from   $Z$ using (\ref{Z}). It also should
  be pointed out that the existence of the solution of the $Z$-RH
   problem (as well as of the $Y$-RH problem) is equivalent to
  the nondegeneracy of the Toeplitz matrix $T_{n}(\phi)$, i.e.
  to the inequality
  \[ D_{n}(\phi) \neq 0,\]
  which we always assume.

\textit{Remark}.  There is a more direct and elegant way to pass
to the $Z$-RH problem which was pointed out by the referee of this
paper.  One first notes that the jump matrix $H$ admits the
factorization,
\[ H(z)=\left(\begin{array}{cc}
z^n & -1 \\
1   &  0\end{array}\right)
\left(\begin{array}{cc}
1 & z^{-n}\phi(z) \\
0 & 1\end{array}\right)
\left(\begin{array}{cc}
z^{-n} & 0\\
-1 & z^n \end{array}\right),
\]
which then suggests the definition
\begin{equation}\label{referee1}
    \tilde Y(z) = \left\{\begin{array}{ll}
          Y(z)\left( \begin{array}{cc}
                z^n & - 1\\
                1   &  0
              \end{array}\right),&\quad |z|<1,\\
                    & \\
          Y(z)\left( \begin{array}{cc}
                z^{n} & 0\\
                1 & z^{-n}
              \end{array}\right) \equiv
            Y(z)\left( \begin{array}{cc}
                1 & 0\\
                z^{-n} & 1
              \end{array}\right)z^{n\sigma_{3}},&\quad |z|>1,
         \end{array}\right. 
  \end{equation}         
  so that the function $\tilde Y$ would satisfy the Riemann-Hilbert problem,
  \begin{itemize}
  \item $\tilde Y$ is holomorphic for all $z\notin C$,
  \item $\tilde{Y}(z)z^{-n\sigma_{3}} \rightarrow I,\quad z\rightarrow \infty$,
  \item $ \tilde{Y}_{-}(z) = \tilde{Y}_{+}(z)
  \left(\begin{array}{cc}
                1 & z^{-n}\phi(z)\\
                0& 1
              \end{array}\right),\quad z \in C$.
  \end{itemize}
   The function $Z(z)$ is related to $\tilde{Y}(z)$ by
    \begin{equation}\label{referee2}
    Z(z) = \sigma_{3} \tilde{Y}(z) \sigma_{3}.
    \end{equation}

  We conclude this section by summarizing the relation of the $Z$-RH problem
  to the orthogonal polynomials on $C$ with respect to the (generally
  complex) weight $\phi$. This relation is due to Deift\cite{bdj1}
  (see also \cite{deift}).\footnote{The $Z$-RH
  problem is the analog for polynomials on the circle of the Riemann-Hilbert
  problem derived in \cite{fik} for polynomials which are orthogonal
  with respect to an exponential weight on the line (see also \cite{bi}).}
  Let $\{P_{k}(z)\}_{k=0, 1,...}$ denote the system
  of the monic polynomials defined  by
  \[
  P_{k}(z) = z^{k} + ...,
  \]
  \[
  \int_{C}P_{n}(z)\bar{P}_{m}(z)\phi(z){dz \over{iz}} = h_{n}\delta_{nm},
  \quad n\geq m,
  \]
  where bar denotes complex conjugation.
  Similarly, introduce a second  system of polynomials, 
  $\{P^{*}_{k}(z)\}_{k=0, 1,...}$,
  by replacing $\phi$ with $\bar\phi$ in   the definition of $P_n$.
   Suppose now that
  \be \label{nonzero}
  D_{k}(\phi) \neq 0, \quad k=1, ..., n+1.
  \ee
  Then (see \cite{Sz1}) both the sets of polynomials
   $\{P_{k}\}_{k=0, 1,..., n}$ and
  $\{P^{*}_{k}\}_{k=0, 1,..., n}$ exist, and the normalization constants
  $h_{k}$ and $h^{*}_{k}$, $k= 1,\ldots,n$,  are all nonzero. In fact, we have
  the explicit representations
  \be \label{orthp3}
  P_{n}(z) = {{D_{n+1}(\phi |z)}\ov {D_{n}(\phi)}},
  \quad h_{n} = 2\pi {{D_{n+1}(\phi)}\ov {D_{n}(\phi)}}, \qquad
  P^{*}_{n}(z) = {{D_{n+1}({\bar{\phi}} |z)}\ov {D_{n}({\bar{\phi}})}},
  \quad h^{*}_{n} = 2\pi {{D_{n+1}({\bar{\phi}})}\ov {D_{n}({\bar{\phi}})}},
  \ee
  where $D_{n+1}(\phi |z)$ denotes the Toeplitz determinant $D_{n+1}(\phi)$
  whose last row is replaced by the row $(1, z, z^{2}, ..., z^{n})$.
  If we define
  \be \label{Q}
  Q_{k} = -{2\pi \over{\bar{h^{*}_{k}}}}\,\bar{P}^{*}_{k}(1/\bar{z})\,z^{k},
  \ee
  and
  \be \label{Z2}
  Z(z)=\left(\begin{array}{cc}
  P_{n}(z) &  {1 \over{2\pi i}}\int_{C}P_{n}(z')(z')^{-n}\phi(z'){dz'
  \over{z'-z}} \\
  {} & {} \\
  Q_{n-1}(z) &  {1 \over{2\pi i}}\int_{C}Q_{n-1}(z')(z')^{-n}\phi(z'){dz'
  \over{z'-z}} \\
  \end{array}\right),
  \ee
  then it is a  calculation to show that this $Z$ defines
  a (unique) solution of the $Z$-RH problem (cf.~\cite{fik}, \cite{bdj1},
  and \cite{bi}). Indeed, the analyticity in $\textbf{C} \setminus {C}$
  and the jump condition follow from the basic properties of Cauchy
  integral, and
  the asymptotic condition at $z = \infty$ is equivalent to the fact
  that the polynomials $P_{n}$ and $P^{*}_{n-1}$ are monic orthogonal
  polynomials with the weights $\phi(z)dz$ and $\bar{\phi}(z)dz$,
  respectively.

  \section{Toeplitz Determinants as  Integrable Systems}
  \setcounter{equation}{0}

  \subsection{Universal recursion relation}

  In this section $\phi$ will be an arbitrary continuous function
  with Fourier coefficients $\phi_{j}$.
  We assume that the associated Toeplitz matrix
   $T_{n}(\phi)$ is invertible. Then the corresponding
  matrix RH problem is uniquely solvable, and the
  following equation connects the Toeplitz determinant
  $D_{n}(\phi)$ with the
  solution $Z$ of the Riemann-Hilbert problem,
  \be \label{recur}
  {D_{n+1} \over{D_{n}}} = Z_{12}(0),
  \ee
  where $Z_{ij}$, $i,j = 1,2$, denotes the entries of matrix $Z$.
  Indeed, using (\ref{orthp3}) we have that
  \[
  {D_{n+1} \over{D_{n}}} = {1 \over{2\pi}}h_{n},
  \]
  On the other hand,  (\ref{Z2}) gives
  \[
  Z_{12}(0) = {1 \over{2\pi}}h_{n},
  \]
  and (\ref{recur}) follows.

  {\it Remark}. One can prove  (\ref{recur}) using only
   the connection with the integrable operator $K_{n}(\phi)$
  introduced in (\ref{inop}). To see this first note
  \ba
  K_{n+1}(z, z') &= &{{(z/z')^{n+1} -1}\ov {z-z'}}\, \,
  {{1-\phi(z')}\ov 2\pi i}
  \\
  &=& {1 \ov{z'}}\sum_{k=0}^{n}\left({z\ov{z'}}\right)^{k}\, \,
  {{1-\phi(z')}\ov 2\pi i}\\
  & =&
  K_{n}(z,z') + f_{1}(z)g_{1}(z'){1\ov{z'}}\, ,
  \ea
  where  $f$  and
  $g$  are defined in (\ref{f}) and
  (\ref{g}), respectively. Attaching to the functions $f$, $g$
  superscript ``$n$'' to denote their $n$ dependence,
   we can rewrite the last equation
  as an operator equation
  \be \label{dk1}
  K_{n+1} = K_{n} + f_{1}^{n}\otimes  g_{1}^{n+1},
  \ee
  where the symbol $a\otimes b$ denotes
  the integral operator with kernel
  $a(z)b(z')$. Recalling the
  definition of $F$,   (\ref{resolvent}),
  it follows from (\ref{dk1}) (cf.~\cite{iiks,tw3}) that
  \ba
  \det(1-K_{n+1})& = & \det(1-K_{n})
  \det(1 - [(1-K_{n})^{-1}f_{1}^{n}]\otimes  g_{1}^{n+1})\\
  &= &\det(1-K_{n})\det(1 - F_{1}^{n}\otimes  g_{1}^{n+1})\\
  &=&\det(1-K_{n})( 1 - \mbox{trace}\, {F_{1}^{n}\otimes  g_{1}^{n+1}})\\
  &=&\det(1-K_{n})\left( 1 - \int_{C}F_{1}(z)g_{1}(z){dz \ov z}\right)
  \ea
  where $F_{1}:= F^{n}_{1}$ and  $g_{1}:= g^{n}_{1}$.
  Thus (see also (\ref{tfred}))
  \be \label{recur2}
  {D_{n+1} \over{D_{n}}} =
  {\det{K_{n+1}} \over{\det K_{n}}} = 1 - \int_{C}F_{1}(z)g_{1}(z){dz \ov z}.
  \ee
  Recalling  (\ref{Y}), we  rewrite (\ref{recur2}) as
  \[
  {D_{n+1} \over{D_{n}}} = Y_{11}(0),
  \]
  which together with (\ref{Z}) yields (\ref{recur}).

  \subsection{Differentiation formulas}

  Here we restrict to the symbol
  \be \label{symbol}
  \phi(z) = e^{tz}\prod_{\alpha = 1}^{M}\left({z-r_{\alpha}
  \ov{z}}\right)^{k_{\alpha}},
  \ee
  where $r_{\alpha} := - p_{i_{\alpha}}$, and we recall (see \S 1)
  that $i_{\alpha} = \max{{\cA}_{\alpha}}$, $k_{\alpha} = |{\cA}_{\alpha}|$,
  and
  \[
  {{\cA}} = \bigcup_{\alpha=1}^{M}{\cA}_{\alpha}
  \]
  is the decomposition of the alphabet ${\cA}$ into subsets
  ${\cA}_{1}, {\cA}_{2}, ..., {\cA}_{M}$ such that $p_{i} = p_{j}$
  if and only if $i$ and $j$ belong the same ${\cA}_{\alpha}$.
  We also recall that
  \[
  \sum_{\alpha=1}^{M} k_{\alpha} = k,
  \]
  and
  \be \label{prob}
  1 >p_{1} \geq p_{2} \geq ... \geq p_{k} > 0,\quad \sum_{j=1}^{k}p_{j} = 1
  \ee
  denote the probabilities assigned to the letters $i = 1, 2,\ldots, k$,
  in the alphabet ${\cA}$. Note that from the probabilistic conditions
  (\ref{prob}) it follows that
  \be \label{rprob1}
  -1 < r_{\alpha} < 0, \quad \alpha = 1,..., M, \quad r_{\alpha} \neq
r_{\beta},
  \, \, \alpha \neq \beta,
  \ee
  and
  \be \label{rprob2}
  \sum_{\alpha = 1}^{M}k_{\alpha}r_{\alpha} = -1.
  \ee

  The symbols $f_{I}$ and   $\phi$ are related by
  \[
  f_{I}(z) = \phi(1/z),
  \]
  and therefore; the corresponding Toeplitz matrices are mutually transpose.
  Thus
  \[
  G_{I}(n;\{p_{i}\}, t) =  D_{n}(\phi).
  \]

  In what follows, we will write  $T_{n}(t)$,
  $K_{n}(t)$ and $D_{n}(t)$ for
  $T_{n}(\phi)$, $K_{n}(\phi)$ and $D_{n}(\phi)$, respectively; or
  $T_{n}(\{p_{i}\}, t)$, $K_{n}(\{p_{i}\}, t)$ and $D_{n}(\{p_{i}\}, t)$
  if the dependence on $p_{1},\ldots, p_{k}$ is of interest.

  \vskip .2in

  We shall derive the differential formulas for the Toeplitz determinant
  $D_{n}(t)$ with respect to the variables $t$ and
  $r_{\alpha}, \alpha = 1, ..., M$ assuming that the latter are subject to
restriction
  (\ref{rprob1}) only, i.e. we only will assume that
  \[
  -1 < r_{\alpha} < 0, \quad \alpha = 1,..., M, \quad r_{\alpha} \neq
r_{\beta},
  \, \, \alpha \neq \beta.
  \]
  The integers $k_{\alpha}$ will be kept constant. This means that {\it when
vary
  the points $r_{\alpha}$ we do not assume restriction (\ref{rprob2}) to
hold\/}.
  We will begin with the $t$ - derivative.

  Since $\partial \phi /\partial{t} = z\phi$,

  \[
  {\partial \ov{\partial{t}}} K_{n}(z,z') =
  {{(z/z')^{n} -1}\ov {z-z'}}\,(-z') \, {\phi(z')\ov 2\pi i}\,
  \]

  \be \label{dif1}
  ={{1-\phi(z')}\ov{2\pi i}} + z{{(z/z')^{n-1} -1}\ov {z-z'}}\, \,
  {{1-\phi(z')}\ov{2\pi i}} - {{(z/z')^{n} -1}\ov {z-z'}}\, \,
  {{z'}\ov{2\pi i}}.
  \ee

  Let $\Lambda$ be the integral operator  with the kernel
  \[
  \Lambda(z,z') = {{(z/z')^{n} -1}\ov {z-z'}}\, \,
  {{z'}\ov{2\pi i}}.
  \]
  Consider the operator product $\Lambda K_{n}$:

  \[
  (\Lambda K_{n})(z,z')=
  {{1-\phi(z')}\ov{2\pi i}}\int_{C}
  {{(z/w)^{n} -1}\ov {z-w}}\, w\,
  {{(w/z')^{n} -1}\ov {w-z'}}\, {dw\ov{2\pi i}}
  \]

  \[
  = {{1-\phi(z')}\ov{2\pi i}}\int_{C}
  \sum_{j,l =0}^{n-1}\left({z\ov{w}}\right)^{l}
  \left({w\ov{z'}}\right)^{j}\, (z')^{-1}\, {dw\ov{2\pi i}}
  = {{1-\phi(z')}\ov{2\pi i}}\sum_{j-l=-1}z^{l}(z')^{-j-1}
  \]

  \[
  = {{1-\phi(z')}\ov{2\pi i}}\sum_{j=0}^{n-2}z^{j+1}(z')^{-j-1}
  ={{1-\phi(z')}\ov{2\pi i}}\, z\, {{(z/z')^{n-1} -1}\ov {z-z'}}.
  \]
  Recalling the definitions of $f$ and $g$, (\ref{f}) and(\ref{g}),
  (\ref{dif1}) can be written compactly as
  \be \label{dk2}
  {\partial \ov{\partial{t}}} K_{n} = -f_{2}\otimes g_{2} - \Lambda (1-K_{n}).
  \ee
  From this formula we see that (cf.~the derivation of (\ref{recur2}))
  \bae
  {\partial \ov{\partial{t}}}\log D_{n}(t) &=& -\mbox{trace}\,
  \left((1-K_{n})^{-1}{\partial \ov{\partial{t}}} K_{n}\right)\nonumber \\
  &=& \mbox{trace}\, F_{2}\otimes g_{2} + \mbox{trace}\, \Lambda \nonumber \\
  &=&\int_{C}F_{2}(z)g_{2}(z)dz,\label{dk3}
  \eae
  where we used the fact that
  \[
  \mbox{trace}\ \Lambda = {n \ov{2\pi i}}\int_{C}dz =0.
  \]
  Recalling  (\ref{Y})
  we convert (\ref{dk3}) into the identity
  \[
  {\partial \ov{\partial{t}}}\log D_{n}(t) = -\mbox{res}_{z=\infty}(Y_{22}(z)),
  \]
  which in terms of the $Z$-function  is
  \[
  {\partial \ov{\partial{t}}}\log D_{n}(t) =
-\mbox{res}_{z=\infty}(z^{n}Z_{22}(z)),
  \]
  or equivalently,
  \be \label{tdk2}
  {\partial \ov{\partial{t}}}\log D_{n}(t) =  (\Gamma_{1})_{22},
  \ee
  where the matrix $\Gamma_{1} = \Gamma_{1}(\{p_{i}\}, t)$ is
  defined by the  expansion,
  \be \label{asymp}
  Z(z) = \left( I + \sum_{j=1}^{\infty}{\Gamma_{j} \ov{z^{j}}}
  \right)\, z^{n\sigma_{3}}, \quad |z| > 1 .
  \ee

  {\it Remark\/}.  In the basis $\{z^{n}\}_{n=-\infty}^{\infty}$,
   (\ref{dk2}) coincides with  (3.22) of \cite{tw2}.

 Equation (\ref{tdk2}) is the $t$ - {\it differentiation formula}, i.e. it
 gives  an expression of the $t$ - derivative
 of $\log D_{n}$ in terms of the solution $Z$ of the Riemann-Hilbert problem.
We shall
 proceed now with the derivation of the $r_{\alpha}$ - differentiation formula.
 Since (we recall that $r_{\alpha}$ are assumed independent and
 that the $k_{\alpha}$
 are kept constant)
 \[
 {\partial \ov{\partial{r_{\alpha}}}}\phi = -{k_{\alpha}
\ov{z-r_{\alpha}}}\phi,
 \]
 the $r_{\alpha}$ -  analog of (\ref{dif1}) reads
\[
{\partial \ov{\partial{r_{\alpha}}}}K_{n}(z,z')  =
k_{\alpha}{{(z/z')^{n} -1}\ov {z-z'}}\, \, {1 \ov{z'-r_{\alpha}}}
\, \, {\phi(z')\ov{2\pi i}}
\]

\be \label{rdif1}
={k_{\alpha}\ov{2\pi i}}{{(z/z')^{n} -1}\ov {z-z'}}\, \,
{1 \ov{z'-r_{\alpha}}}\, \, - \, \,
k_{\alpha}{{(z/z')^{n} -1}\ov {z-z'}}\, \, {1 \ov{z'-r_{\alpha}}}
\, \, {{1-\phi(z')}\ov{2\pi i}} .
\ee

Introducing the integral operator $\Lambda_{\alpha}$ with the kernel,
\[
\Lambda_{\alpha}(z,z') = {k_{\alpha}\ov{2\pi i}}{{(z/z')^{n} -1}\ov {z-z'}}
\, \,{1 \ov{z'-r_{\alpha}}},
\]
we consider again the operator product $\Lambda_{\alpha}K_{n}$. The residue
type
calculations, similar to the ones used in the $t$ - case, yield the equation
\[
(\Lambda_{\alpha}K_{n})(z,z') =
{k_{\alpha}\ov{2\pi i}}(1-\phi(z'))
\left [{{(z/z')^{n} -1}\ov {(z'-z)(r_{\alpha} -z)}} +
{{(r_{\alpha}/z')^{n} -1}\ov {(r_{\alpha} -z)(r_{\alpha} -z')}}\right ],
\]
which in turn impies that (\ref{rdif1}) can be rewritten as
\[
{\partial \ov{\partial{r_{\alpha}}}}K_{n}(z,z')=
{k_{\alpha}\ov{2\pi i}}{{1-\phi(z')}\ov{(r_{\alpha} -z)(r_{\alpha} -z')}}
\left [\left ({r_{\alpha}\ov z'}\right )^{n} -  \left ({z\ov z'}\right
)^{n}\right ]
+ [\Lambda_{\alpha}(1-K_{n})](z,z').
\]

With the help of the vector functions,
\[
\tilde{f}(z):= {1\ov{z-r_{\alpha}}}f(z), \quad
\tilde{g}(z):= {1\ov{z-r_{\alpha}}}g(z),
\]
the last equation can be  transformed into  the following compact form (cf.
(\ref{dk2}))
\be \label{rdk2}
{\partial \ov{\partial{r_{\alpha}}}} K_{n}
= k_{\alpha}r^{n}_{\alpha}\tilde{f}_{2}\otimes \tilde{g}_{1} -
k_{\alpha}\tilde{f}_{1}\otimes \tilde{g}_{1}
 + \Lambda_{\alpha} (1-K_{n}).
\ee
Let the vector function $\tilde{F}(z) = (\tilde{F}_{1}(z),
\tilde{F}_{2}(z))^T$ be defined by the equation
\[
\tilde{F}_{j}:= (1-K_{n})^{-1}\tilde{f}_{j}, \quad j = 1,2.
\]
We observe that
\be \label{tildeF}
\tilde{F}(z) = {1\ov{z-r_{\alpha}}}Y^{-1}(r_{\alpha})F(z),
\ee
where the matrix function $Y(z)$ is the solution of the $Y$-RH 
problem corresponding to $D_{n}(t)$. Indeed by the definition
of the vector  function $F(z)$ (see (\ref{resolvent})) its
component $F_{j}(z)$  satisfies the integral equation,

\be \label{intF}
F_{j}(z) - \int_{C}K_{n}(z,z')F_{j}(z')dz' = f_{j}(z).
\ee
Dividing both sides of this equation by $(z - r_{\alpha})$,
using the formula
\[
K_{n}(z,z') = {{f^{T}(z)g(z')}\ov{z-z'}},
\]
and  simple algebra,
we can rewrite (\ref{intF}) as an equation for the ratio
$F_{j}/(z-r_{\alpha})$:
\[
\left[ {F_{j}(z)\ov{z-r_{\alpha}}}\right ] -
\int_{C}K_{n}(z,z')\left[ {F_{j}(z')\ov{z'-r_{\alpha}}}\right ]dz'
+ \int_{C}{\tilde{f}^{T}(z)g(z')}\left[ {F_{j}(z')\ov{z'-r_{\alpha}}}\right
]dz'
=\tilde{f}_{j}(z).
\]
By applying  the operator $(1-K_{n})^{-1}$
to the both sides of this equation it can be transformed into the
relation,
\[
\left[ {F_{j}(z)\ov{z-r_{\alpha}}}\right ] +
\int_{C}{\tilde{F}^{T}(z)g(z')}\left[ {F_{j}(z')\ov{z'-r_{\alpha}}}\right ]dz'
=\tilde{F}_{j}(z),
\]
or
\[
{1\ov{z-r_{\alpha}}}F_{j}(z) + \sum_{i=1}^{2}\tilde{F}_{i}(z)
\int_{C}g_{i}(z')F_{j}(z'){dz'\ov {z'-r_{\alpha}}}=\tilde{F}_{j}(z).
\]

The last equation in turn can be viewed as the linear algebraic system for
the vector $\tilde{F}(z)$,
\be \label{eqtF}
\tilde{F}_{j}(z) - \sum_{i=1}^{2}A_{ji}\tilde{F}_{i}(z) =
{1\ov{z-r_{\alpha}}}F_{j}(z), \quad j=1,2,
\ee
where the matrix $A$ is given by the formula,
\[
A_{ji} = \int_{C}F_{j}(z')g_{i}(z'){dz'\ov {z'-r_{\alpha}}}.
\]
Equation (\ref{tildeF}) follows directly from (\ref{eqtF}) in virtue of
definition (\ref{Y}) of the matrix function $Y(z)$.

We  now able to finish the derivation of the $r_{\alpha}$ -
differentiation formula for the Toeplitz determinant $D_{n}(t)$.
In fact from (\ref{rdk2})  it follows that
(cf.~the derivation of (\ref{dk3}))
\bae
{\partial \ov{\partial{r_{\alpha}}}}\log D_{n}(t) &=& -\mbox{trace}\,
\left((1-K_{n})^{-1}{\partial \ov{\partial{r_{\alpha}}}} K_{n}\right)\nonumber
\\
&=&- k_{\alpha}r^{n}_{\alpha}\mbox{trace}\, \tilde{F}_{2}\otimes \tilde{g}_{1}
 + k_{\alpha}\mbox{trace}\, \tilde{F}_{1}\otimes \tilde{g}_{1} -
\mbox{trace}\,\Lambda_{\alpha} \nonumber \\
&=& -k_{\alpha}r^{n}_{\alpha}\int_{C}\tilde{F}_{2}(z)g_{1}(z){dz\ov
{z-r_{\alpha}}}
+ k_{\alpha}\int_{C}\tilde{F}_{1}(z)g_{1}(z){dz\ov{z-r_{\alpha}}},\label{rdk3}
\eae
where, similar to the $t$ - derivative case, we used the fact that
\[
  \mbox{trace}\ \Lambda_{\alpha} = {nk_{\alpha} \ov{2\pi i}}\int_{C}
{dz\ov{z(z-r_{\alpha})}} =0.
\]
Using now (\ref{tildeF}) and the fact that $\det Y(z) \equiv 1$ we derive
from (\ref{rdk3}) that
\ba
{\partial \ov{\partial{r_{\alpha}}}}\log D_{n}(t) &=&
k_{\alpha}r^{n}_{\alpha}Y_{21}(r_{\alpha})
\int_{C}F_{1}(z)g_{1}(z){dz\ov{(z-r_{\alpha})^2}}
-k_{\alpha}r^{n}_{\alpha}Y_{11}(r_{\alpha})
\int_{C}F_{2}(z)g_{1}(z){dz\ov{(z-r_{\alpha})^2}} \nonumber \\
&&\; +k_{\alpha}Y_{22}(r_{\alpha})
\int_{C}F_{1}(z)g_{1}(z){dz\ov{(z-r_{\alpha})^2}}
-k_{\alpha}Y_{12}(r_{\alpha})
\int_{C}F_{2}(z)g_{1}(z){dz\ov{(z-r_{\alpha})^2}}\, \, ,
\ea
or
\bae
{\partial \ov{\partial{r_{\alpha}}}}\log D_{n}(t)
&=& -k_{\alpha}r^{n}_{\alpha}Y_{21}(r_{\alpha})Y'_{11}(r_{\alpha})
+k_{\alpha}r^{n}_{\alpha}Y_{11}(r_{\alpha})Y'_{21}(r_{\alpha}) \nonumber \\
&&\; -\, k_{\alpha}Y_{22}(r_{\alpha})Y'_{11}(r_{\alpha})
+k_{\alpha}Y_{12}(r_{\alpha})Y'_{21}(r_{\alpha}),\label{rdk4}
\eae
where we use the notation,
\[
Y'_{ij}(r_{\alpha}) := {\partial{Y_{ij}(z)} \ov{\partial{z}}}|_{z=r_{\alpha}},
 \quad i,j = 1,2.
\]
Equation (\ref{rdk4}) can be also rewritten as
\bae
{\partial \ov{\partial{r_{\alpha}}}}\log D_{n}(t)
&= & -k_{\alpha}\Big (r^{n}_{\alpha}Y_{21}(r_{\alpha})+
Y_{22}(r_{\alpha})\Big )Y'_{11}(r_{\alpha}) \nonumber \\
&& \;+ k_{\alpha}\Big (r^{n}_{\alpha}Y_{11}(r_{\alpha})+
Y_{12}(r_{\alpha})\Big )Y'_{21}(r_{\alpha}),\label{rdk5}
\eae
which in turn can be  transformed into an expression of
the $\partial{\log D_{n}(t)}/\partial{r_{\alpha}}$ in terms
of the $Z$-function. Indeed recalling formulae (\ref{Z}),
(\ref{F}), and (\ref{f}), we see that inside the unit
circle $C$ the following equation takes place,
\[
Z(z) = \left( \begin{array}{cc}
  z^{n}Y_{11}(z) + Y_{12}(z) & Y_{11}(z)\\
  -z^{n}Y_{21}(z) - Y_{22}(z) & -Y_{21}(z)
  \end{array}\right) , \quad |z|< 1 ,
\]
so that (\ref{rdk5}) can be converted into the $r_{\alpha}$-differentiation 
formula,
\be \label{rdk6}
{\partial \ov{\partial{r_{\alpha}}}}\log D_{n}(t)
=- k_{\alpha}\Big (Z_{11}(r_{\alpha})Z'_{22}(r_{\alpha})
- Z_{21}(r_{\alpha})Z'_{12}(r_{\alpha})\Big ).
\ee
  \subsection{Schlesinger equations}
  In this section we show that  $D_{n}(t)$
  is the Jimbo-Miwa-Ueno $\,\tau$-function of the generalized Schlesinger
  system of nonlinear differential equations describing the
  isomonodromy deformations of the
  $2\times 2$ matrix linear ODE which has $M+1$ simple poles in
  the finite complex plane and one Poincar\'e index 1 irregular
  singular point at infinity. We will also evaluate the relevant monodromy
  data that single out the $D_{n}(t)$ from all the other solutions of the
  Schlesinger system.
  In the uniform
  case, when all $p_{i}$ are equal, the system reduces
  to the particular case of Painlev\'e V equation, i.e. we are back
  to the uniform result of \cite{tw2}.

  Define
  \be \label{Phi0}
  \Phi^{0}(z) = e^{{tz \ov{2}}\sigma_{3}}
  \left(\begin{array}{cc}
  1 & 0 \\
  0 & z^{n}{\psi}^{-1}(z)
  \end{array}\right)
  \ee
  where
  \be \label{psi}
  \psi(z) = \prod_{\alpha = 1}^{M}\left({z-r_{\alpha}
  \ov{z}}\right)^{k_{\alpha}}.
  \ee
  We note that that the product
  \[
  e^{tz}\psi(z) := \phi(z)
  \]
  is our symbol, i.e.\ the function defined in
  (\ref{symbol}). We also note that  $\Phi^{0}$ is
  analytic and invertible in $\textbf{C} \setminus \{0, r_{1}, ..., r_{M}\}$,
  and that it satisfies the linear differential equations
  \bae
  \Phi^{0}_{z}(z)&=&
  \Omega(z) \,\Phi^{0}(z), \label{eqz0}\\
  \Phi^{0}_{t}(z)&=&
  {z \ov 2}\,\sigma_{3}\, \Phi^{0}(z),\label{eqt0}\\
  \Phi^{0}_{r_{\alpha}}(z)&=&
  {k_{\alpha} \ov{z-r_{\alpha}}}
  \left(\begin{array}{cc}
  0 & 0 \\
  0 & 1 \end{array}\right)
  \Phi^{0}(z),\label{eqr0}
  \eae
  where $\Omega$ is the rational matrix function
  \be \label{omega}
  \Omega(z) = {t \ov 2}\,\sigma_{3}
  + {n+k \ov z}\left(\begin{array}{cc}
  0 & 0 \\
  0 & 1 \end{array}\right)
  - \sum_{\alpha = 1}^{M}
  {k_{\alpha} \ov {z-r_{\alpha}}} \left(\begin{array}{cc}
  0 & 0 \\
  0 & 1 \end{array}\right).
  \ee
  (Subscripts on $\Phi^0$ denote differentiation.)
  Introduce
  \be \label{Phi}
  \Phi(z) = Z(z)\Phi^{0}(z),
  \ee
  where $Z$ is the solution of the $Z$-RH problem corresponding
  to our symbol $\phi$, and consider the logarithmic derivative
  \be \label{Bdef}
  B(z) := \Phi_{z}(z){\Phi}^{-1}(z).
  \ee
  The key observation is that  $B$ is
  continuous across the contour $C$. Indeed, the $Z$-jump
  matrix $S$ (see (\ref{Sz})) admits the following factorization,
  \be
  S(z) = \Phi^{0}(z)
  \left(\begin{array}{cc}
  1 & 1 \\
  0 & 1 \end{array}\right)
  (\Phi^{0}(z))^{-1},
  \label{factNew}\ee
  so that the $\Phi$-jump matrix
  does not depend on $z$. In fact we have
  \be \label{Phijump}
  \Phi_{+}(z) = \Phi_{-}(z)
  \left(\begin{array}{cc}
  1 & 1 \\
  0 & 1 \end{array}\right), \quad z \in C.
  \ee
  This implies that
  \[
  B_{+}(z) = B_{-}(z), \quad z \in C,
  \]
  and hence the function $B(z)$ is an analytic function
  on $\textbf{C} \setminus \{ 0, r_{1}, ..., r_{M}\}$. We
  also recall that the only conditions which we impose
  on the points $r_{\alpha}$ are the inequalities (\ref{rprob1}),
  i.e.,
  \be \label{rprob}
  -1 < r_{\alpha} < 0, \quad \alpha = 1,..., M, \quad r_{\alpha} \neq
r_{\beta},
  \, \, \alpha \neq \beta.
  \ee

  We now calculate the principal part of
  $B$ at each of its singular points. Since $Z$ is
  holomorphic and invertible inside of $C$,
   it follows from
  \be \label{B}
  B(z) = Z(z)\Omega(z)Z^{-1}(z) + Z_{z}(z)Z^{-1}(z)
  \ee
   that in a neighborhood
  of $z = r_{\alpha}$,
  \be \label{r}
  B(z) = -{k_{\alpha}\ov{z-r_{\alpha}}}
  Z(r_{\alpha})\left(\begin{array}{cc}
  0 & 0 \\
  0 & 1 \end{array}\right)Z^{-1}(r_{\alpha})
  + \sum_{j=0}^{\infty}b^{\alpha}_{j}(z-r_{\alpha})^{j}.
  \ee
  Likewise in a neighborhood of $z=0$,
     (\ref{B}) and (\ref{omega}) imply that
  \be \label{0}
  B(z) = {n+k\ov z}
  Z(0)\left(\begin{array}{cc}
  0 & 0 \\
  0 & 1 \end{array}\right)Z^{-1}(0)
  + \sum_{j=0}^{\infty}b^{0}_{j}z^{j}.
  \ee
  Finally, from (\ref{B}) and the Laurent expansion
  (\ref{asymp}) we obtain the power series of $B$ at $\infty$,
  \be \label{infty}
  B(z) = {t \ov 2}\sigma_{3} + \sum_{j=1}^{\infty}b^{\infty}_{j}z^{-j}.
  \ee

  Equations (\ref{r})--(\ref{infty}) imply that $B$
  is a rational function,
  \be \label{Bequ}
  B(z) =  {t \ov 2}\sigma_{3} +{B_{0}\ov z} + \sum_{\alpha = 1}^{M}
  {B_{\alpha}\ov {z-r_{\alpha}}},
  \ee
  with the matrix residues given by
  \be \label{B0}
  B_{0} = Z(0)\left(\begin{array}{cc}
  0 & 0 \\
  0 & n+k \end{array}\right)Z^{-1}(0),
  \ee
  \be \label{Ba}
  B_{\alpha} =
  -Z(r_{\alpha})\left(\begin{array}{cc}
  0 & 0 \\
  0 & k_{\alpha} \end{array}\right)Z^{-1}(r_{\alpha}),\quad \alpha = 1,\ldots,M.
  \ee
  Thus from  (\ref{Bdef}) we conclude that
  $\Phi$ satisfies the linear differential equation,
  \be \label{Phiequ_z}
  \Phi_{z}(z) = B(z)\Phi(z),
  \ee
  with the coefficient matrix $B$ determined by
  (\ref{Bequ})--(\ref{Ba}).

  {\it Remark\/}. In soliton theory (see \cite{nmpz}), the method
  that we used to derive
    (\ref{Phiequ_z}) is called
  \textit{Zakharov-Shabat dressing}
  of the vacuum equation (\ref{eqz0}).  We also note
that, as is common in the analysis of soliton equations, we
have moved the exponential factor $e^{\pm z/2}$ to the asymptotic
condition at $z=\iy$.

  Let us now dress the $t$-vacuum equation (\ref{eqt0}), i.e.\
  consider the $t$-logarithmic  derivative of  $\Phi$
  \be \label{Vdef}
  V(z) := \Phi_{t}(z){\Phi}^{-1}(z).
  \ee
  The $\Phi$-jump matrix (\ref{Phijump}) does not depend on $t$ as well.
  Hence
  \[
  V_{+}(z) = V_{-}(z), \quad z \in C,
  \]
  and  $V$ is analytic  on
  $\textbf{C} \setminus \{0, r_{1}, ..., r_{m}\} $. In fact,
  since
  \be \label{V}
  V(z) = Z(z){z\ov 2}\sigma_{3}Z^{-1}(z) + Z_{t}(z)Z^{-1}(z),
  \ee
  (cf. (\ref{B})) and $Z$ is holomorphic at  the
   points $\{0, r_{1}, ..., r_{M}\}$, we conclude that
   $V$ is entire. Moreover, from the expansion (\ref{asymp})
  we have that
  \[
  V(z) = {z\ov 2}\sigma_{3} + {1\ov 2}[\sigma_{3}, \Gamma_{1}]
  + \sum_{j=1}^{\infty}v_{j}z^{-j},\quad |z| > 1,
  \]
  and hence
  \be \label{Vequ}
  V(z) = {z\ov 2}\sigma_{3} + {1\ov 2}[\sigma_{3}, \Gamma_{1}].
  \ee
  ($[L,M] := LM-ML$.)

  This in turn yields the $t$-equation for  $\Phi$,
  \be \label{Phiequ_t}
  \Phi_{t}(z) = V(z)\Phi(z),
  \ee
  where the coefficient matrix $V$ is defined by the equations,
  \bae
  V(z) &=& {z\ov 2}\sigma_{3} + V_{0},\label{Vequ1}\\
  V_{0} &=&{1\ov 2}[\sigma_{3}, \Gamma_{1}],\label{V0}\\
  \Gamma_{1}&=& -\mbox{res}_{z=\infty}(Z(z)z^{-n\sigma_{3}})
  \>\> (\textrm{see also (\ref{asymp})}).\label{G1}
  \eae

  Equations (\ref{Phiequ_z}) and (\ref{Phiequ_t}) form an
  overdetermined system for the function $\Phi$  in the
  variables $z$ and $t$. From
  the compatibility condition,
  \[
  \Phi_{zt} = \Phi_{tz},
  \]
  we derive the following equation for the coefficient matrices
  $B$ and $V$,
  \be \label{schles1}
  B_{t}(z) - V_{z}(z) = [V(z), B(z)],
  \ee
  or, taking into account (\ref{Vequ1}),
  \be \label{schles2}
  B_{t}(z) - {\sigma_{3} \ov 2} = {z\ov 2}[\sigma_{3}, B(z)]
  + [V_{0}, B(z)].
  \ee
  Since this equation is satisfied identically in $z$,
  a comparison of the principal parts of both the sides at
  $z = 0, r_{1},\ldots, r_{M}$,  then leads to  the differential
  relations,
  \be \label{schles3}
  {\partial{B_{\alpha}} \ov \partial{t}} =
  [{r_{\alpha}\ov 2}\sigma_{3} + V_{0},
  B_{\alpha}].
  \quad \alpha = 0, 1, ..., M,
  \ee
  (It is notationally convenient to define
  $r_{0}= 0$.)

  The important point now is that the matrix $V_{0}$
  can be expressed in terms of the matrices $B_{\alpha}$, so that
  relations (\ref{schles3}) form a closed system of nonlinear
  ODEs for the matrix residues $B_{\alpha}$. In fact, expanding
  both sides of  (\ref{B}) in a Laurent series at
  $z=\infty$,  using  (\ref{omega}), (\ref{Bequ}),
  and (\ref{asymp}), and equating the terms of order $z^{-1}$ we
  have
  \bae
  \sum_{\alpha=0}^{m} B_{\alpha} &= &{t\ov 2}\,[\Gamma_{1}, \sigma_{3}]
  + (n+k)\left(\begin{array}{cc}
  0 & 0 \\
  0 & 1 \end{array}\right)
  -\sum_{\alpha=1}^{m}k_{\alpha}
  \left(\begin{array}{cc}
  0 & 0 \\
  0 & 1 \end{array}\right)
  + n\,\sigma_{3}\nonumber\\
  &= &{t\ov 2}\,[\Gamma_{1}, \sigma_{3}] + n
  \left(\begin{array}{cc}
  0 & 0 \\
  0 & 1 \end{array}\right)
  + n\,\sigma_{3} \nonumber\\
  &= &{t\ov 2}\,[\Gamma_{1}, \sigma_{3}] + {n\ov 2}\,\sigma_{3} + {n\ov 2}\,I.
   \label{trace1}
  \eae
  Comparing the last equation with (\ref{V0}) we obtain
  \be \label{V0B}
  V_{0} = {1\ov t} \sum_{\alpha=0}^{m} B_{\alpha}
  -{n\ov 2t}\sigma_{3} - {n\ov 2t}I,
  \ee
  so that  (\ref{schles3}) becomes
  \be \label{schles4}
  {\partial{B_{\alpha}} \ov \partial{t}} = {{n-tr_{\alpha}}\ov 2t}[
  B_{\alpha},\sigma_{3}]
   + \sum_{\gamma =0}^{m}{[B_{\gamma}, B_{\alpha}]\ov t},
  \quad \alpha = 0, 1,\ldots, M.
  \ee

  If we vary the points $r_{\alpha}$,
  then we  obtain $M$ additional linear differential equations for
   $\Phi(z)= \Phi(z,t, r_{1}, ..., r_{M})$,
  \be \label{rder}
  \Phi_{r_{\alpha}}(z) = -{{B_{\alpha}}\ov{z-r_{\alpha}}}\Phi(z),\quad
  \alpha = 1, ..., M.
  \ee
  Indeed, introducing the $r_{\alpha}$ - logarithmic derivative,
  \[
  U_{\alpha}(z) := \Phi_{r_{\alpha}}(z){\Phi}^{-1}(z),
  \]
  and using exactly the same line of arguments as before,
  we conclude that $U_{\alpha}$ is analytic  on
  $\textbf{C} \setminus \{0, r_{1}, ..., r_{m}\} $. Simultaneously,
  the $r_{\alpha}$ - vacuum equation (\ref{eqr0}) implies the
  identity,
  \be \label{U}
  U_{\alpha}(z) =
 {k_{\alpha}\ov{z-r_{\alpha}}}
  Z(z)\left(\begin{array}{cc}
  0 & 0 \\
  0 & 1 \end{array}\right)
  Z^{-1}(z) + Z_{r_{\alpha}}(z)Z^{-1}(z),
  \ee
  (cf. (\ref{B}) and (\ref{V})) which indicates that the
  only singularity of $U_{\alpha}$ is a simple pole at
  $z=r_{\alpha}$ with
  \[
  {k_{\alpha}\ov{z-r_{\alpha}}}
  Z(r_{\alpha})\left(\begin{array}{cc}
  0 & 0 \\
  0 & 1 \end{array}\right)
  Z^{-1}(r_{\alpha})
  \equiv -{{B_{\alpha}}\ov{z-r_{\alpha}}}
  \quad (\mbox{see also (\ref{Ba})})
  \]
  as the corresponding principal part.
  Moreover, taking into account that the
  asymptotics of $Z(z)$ as $z \rightarrow \infty$ does not
  depend on $r_{\alpha}$ we conclude that
  \[
  U_{\alpha}(z) \rightarrow 0, \quad z \rightarrow \infty,
  \]
  and hence
  \[
   U_{\alpha}(z) =  -{{B_{\alpha}}\ov{z-r_{\alpha}}}.
  \]
  Equation (\ref{rder}) now follows.

  The compatibility conditions of equations (\ref{rder}) with
 (\ref{Phiequ_z}) lead to the nonlinear $r$-differential
  equations for the matrices
  $B_{\alpha} = B_{\alpha}(t, r_{1}, ..., r_{M})$,
  \bae
  {{{\pl}B_{\alpha}}\ov {{\pl}r_{\gamma}}}
  &=& {[B_{\alpha}, B_{\gamma}]\ov{r_{\alpha} - r_{\gamma}}},
  \quad \alpha \neq \gamma = 1,\ldots, M,\label{schles5a}\\
  {{{\pl}B_{0}}\ov {{\pl}r_{\alpha}}}
  &=& {[B_{0}, B_{\alpha}]\ov{r_0 - r_{\alpha}}},
  \quad \alpha = 1,\ldots, M,\label{schles5b}\\
  {{{\pl}B_{\alpha}}\ov {{\pl}r_{\alpha}}}
  &=& \sum_{\gamma \neq \alpha}{[B_{\gamma}, B_{\alpha}]\ov{r_{\gamma} -
  r_{\alpha}}},
  \quad \alpha = 1,\ldots, M.\label{schles5c}
  \eae
  which supplement $t$-equation (\ref{schles4}).

  The total system
  ((\ref{schles4}), (\ref{schles5a})--(\ref{schles5c}))
  of nonlinear PDEs is the (generalized) system of Schlesinger
  equations which describes the isomonodromy deformations
  (see e.g.~\cite{jmu, jimbo}) of
  the coefficients of the $2\times 2$ system of linear ODEs having $M+1$
  regular singularities at the points $ z = r_{\alpha}, \alpha = 0, ..., M$
  and an irregualr singular point of Poincar\'e index 1 at
  infinity (see (\ref{Phiequ_z}),
  (\ref{Bequ})),
  \be \label{fucsh}
  {{d\Phi(z)}\ov dz} = B(z)\Phi(z), \quad
  B(z) =  {t \ov 2}\sigma_{3}  + \sum_{\alpha = 0}^{M}
  {B_{\alpha}\ov {z-r_{\alpha}}}.
  \ee
  The monodromy data of equation (\ref{fucsh}) which single out
  the solution of ((\ref{schles4}), (\ref{schles5a})--(\ref{schles5c})),
  which we are interested in, coincide,
  after the proper normalization, with the data of the
  $Z$-RH problem. More precisely, let us denote
  $\Phi^{\infty}(z)$ the analytic continuation of $\Phi(z)$ from
  $|z|>1$ to the whole  complex $z$-plane. Then, the $Z$-RH
  problem and equation (\ref{Phi}) imply the following representations
  of the function
  $\Phi^{\infty}(z)$ in the neighborhoods of its singular points,
  \bae
  \Phi^{\infty}(z) &=& \hat{\Phi}_{\alpha}(z)
  \left(\begin{array}{cc}
  1 & 0\\
  0 & (z-r_{\alpha})^{-k_{\alpha}}
  \end{array}\right)
  \left(\begin{array}{cc}
  1 & \hspace{-1.2ex} -1\\
  0 & 1
  \end{array}\right),\quad z\in U_{ r_{\alpha}},\label{ralpha}\\
  \Phi^{\infty}(z) &=& \hat{\Phi}_{0}(z)
  \left(\begin{array}{cc}
  1 & 0\\
  0 & z^{n+k}
  \end{array}\right)
  \left(\begin{array}{cc}
  1 & \hspace{-1.2ex} -1\\
  0 & 1
  \end{array}\right),\quad z\in U_{0},\label{r0}\\
  \Phi^{\infty}(z) &=& \hat{\Phi}_{\infty}(z)e^{{{tz}\ov 2}\sigma_{3}}
  \left(\begin{array}{cc}
  z^{n} & 0\\
  0 & 1
  \end{array}\right),\quad z\in U_{\infty}, \quad
  \hat{\Phi}_{\infty}(\infty)= I. \label{rinfty}
  \eae
  Here $\hat{\Phi}_{\alpha}(z)$, $\hat{\Phi}_{0}(z)$,
  and $\hat{\Phi}_{\infty}(z)$ denote the matrix functions
  which are holomorphic and invertible in the
  neighborhoods $U_{ r_{\alpha}}$, $U_{0}$, and $U_{\infty}$
  respectively.
  Formulae (\ref{ralpha})--(\ref{rinfty})
   allow us to identify the diagonal matrices,
  \be \label{fmon}
  E_{\alpha} =\left(\begin{array}{cc}
  0 & 0\\
  0 & -k_{\alpha}
  \end{array}\right), \quad
  E_{0}=\left(\begin{array}{cc}
  0 & 0\\
  0 & n+k
  \end{array}\right), \quad \mbox{and} \quad
  E_{\infty}=\left(\begin{array}{cc}
  \hspace{-1.4ex}-n & 0\\
  0 & 0
  \end{array}\right),
  \ee
  as the formal monodromy exponents
  (cf.~\cite{jmu, jimbo})  of $\Phi^{\infty}(z)$
  at the points $r_{\alpha}$, $0$, and $\infty$
  respectively. The corresponding connection matrices, i.e.
  the matrices $C_{\alpha}$ in the representations,
  \[
  \Phi^{\infty}(z) = \hat{\Phi}_{\alpha}(z)
  (z-r_{\alpha})^{E_{\alpha}}C_{\alpha}, \quad \alpha = 0, ..., M,
  \]
  all are given by
  \be \label{connect}
  C_{\alpha} = \left(\begin{array}{cc}
  1 &\hspace{-1.2ex} -1\\
  0 & 1
  \end{array}\right), \quad \alpha = 0,\ldots, M.
  \ee

  Since the numbers $k_{\alpha}$, $k$,
  and $n$ are  integers, all the monodromy matrices of
  $\Phi^{\infty}(z)$ are trivial. There are also no Stokes' matrices
  at the irregular singular point $z = \infty$ since the asymptotic
  series (\ref{asymp}), as a Laurent series, converges in a disk centered at
  infinity. Therefore the complete monodromy data
  of the linear system (\ref{fucsh})
  for our random word problem,  consists of (\textit{i}) (\ref{fmon}),
  the formal monodromy exponents at the singular points, and (\textit{ii})
   (\ref{connect}),  the corresponding connection matrices.

  \subsection{Toeplitz determinant as a $\tau$-function}

  In this section we shall derive the exact formulae for the logarithmic
derivatives
  of the Toeplitz determinant $D_{n}(t, r_{1}, ..., r_{M})$ in terms of the
  matrices $B_{\alpha}$ which,
  as we saw in the previous section, satisfy the Schlesinger equations
  ((\ref{schles4}), (\ref{schles5a})--(\ref{schles5c})).
  To this end we will exploit  (\ref{tdk2}) and (\ref{rdk6}) whose right hand
sides
  we will express via $B_{\alpha}$ using a technique  similar to the one
  that led to (\ref{V0B}). We begin with (\ref{tdk2}).

  Equation (\ref{V0B}) was obtained by expanding both sides of (\ref{B})
  about $\iy$ and  then equating the terms of order $z^{-1}$.
  Let us now analyze the terms of order $z^{-2}$.  From
  (\ref{omega}) it follows that
  \ba
  \Omega(z) &=& {t\ov 2}\,\sigma_{3} +{n\ov z}\,
  \left(\begin{array}{cc}
  0 & 0 \\
  0 & 1 \end{array}\right)
  - {1\ov z^{2}}\,\sum_{\alpha = 1}^{M}r_{\alpha}k_{\alpha}
  \,\left(\begin{array}{cc}
  0 & 0 \\
  0 & 1 \end{array}\right)
  + O\left({1\ov z^{3}}\right)\\
  &:=&   {t\ov 2}\sigma_{3} + {\Omega_{1} \ov z} +
  {\Omega_{2} \ov z^{2}}  + O\left({1\ov z^{3}}\right).
  \ea
  Combining this  with expansion (\ref{asymp})
  of $Z$, we get the following expression
  for the order $z^{-2}$ term of the right hand side of (\ref{B}):
  \[ {t\ov 2}\,[\sigma_{3}, \Gamma_{1}]\,\Gamma_{1}
  + {t\ov 2}\,[\Gamma_{2}, \sigma_{3}] + [\Gamma_{1}, \Omega_{1}]
  + \Omega_{2} - \Gamma_{1} + {n\ov 2}\,[\Gamma_{1}, \sigma_{3}].
  \]
  The  order $z^{-2}$ term of the left hand side follows
  directly from (\ref{Bequ}):
  \[ \sum_{\alpha = 1}^{M}r_{\alpha}B_{\alpha}.\]
  Equating the two expressions we arrive at
  \be \label{trace2}
  \sum_{\alpha = 1}^{M}r_{\alpha}B_{\alpha}
  = {t\ov 2}\,[\sigma_{3}, \Gamma_{1}]\,\Gamma_{1}
  + {t\ov 2}\,[\Gamma_{2}, \sigma_{3}] + [\Gamma_{1}, \Omega_{1}]
  + \Omega_{2} - \Gamma_{1} + {n\ov 2}\,[\Gamma_{1}, \sigma_{3}].
  \ee
  This equation together with  (\ref{trace1})
  determines  $\Gamma_{1}$ in terms
  of the matrices $B_{\alpha}$. Indeed,  (\ref{trace1})
  gives the off diagonal part of $\Gamma_{1}$. Using
  that for $L$ diagonal
  \[\mbox{diag}\, [P, L] = 0, \]
  we have from  (\ref{trace2}) that
  \be \label{Gammadiag1}
  \mbox{diag}\, \Gamma_{1} =
  - \mbox{diag}\,\sum_{\alpha = 1}^{m}r_{\alpha}B_{\alpha} +
  {t\ov 2}\,\mbox{diag}\,([\sigma_{3}, \Gamma_{1}]\Gamma_{1})
  + \Omega_{2}.
  \ee
  Using the identity that for any $2\times 2$ matrix $P$,
  \[
  \mbox{diag}\,([\sigma_{3}, P]P) = - {1\ov 2}[\sigma_{3}, P]^{2}
  \sigma_{3},
  \]
  we obtain from (\ref{Gammadiag1}) and (\ref{trace1})
  the final expression for the diagonal part of $\Gamma_{1}$,
  \bae
  \mbox{diag}\, \Gamma_{1} &=&
  - \mbox{diag}\,\sum_{\alpha = 1}^{M}r_{\alpha}B_{\alpha}
  -{1\ov t}\,\left(\sum_{\alpha = 0}^{M}B_{\alpha} -{n\ov 2}\sigma_{3}
  - {n\ov 2}I\right)
  \left(\sum_{\alpha = 0}^{M}B_{\alpha} -{n\ov 2}\sigma_{3}
  - {n\ov 2}I\right)\sigma_{3} + \Omega_{2}\nonumber\\
  &=& - \mbox{diag}\,\sum_{\alpha = 1}^{M}r_{\alpha}B_{\alpha}
  -{1\ov t}\,\mbox{diag}\,\left(\left(\sum_{\alpha = 0}^{M}
  B_{\alpha} -{n\ov 2}\sigma_{3}
  - {n\ov 2}I\right)\sum_{\alpha = 0}^{M}B_{\alpha}\sigma_{3}\right)
   + \Omega_{2}\nonumber\\
  &=& - \mbox{diag}\,\sum_{\alpha = 1}^{M}r_{\alpha}B_{\alpha}
  -{1\ov t}\mbox{diag}\,\left(\sum_{\alpha, \gamma = 0}^{M}
  B_{\alpha}B_{\gamma}\sigma_{3} -{n\ov 2}
  \sum_{\alpha = 0}^{M}(\sigma_{3} + I)B_{\alpha}\sigma_{3}\right)
   + \Omega_{2}\nonumber\\
  &=& - \mbox{diag}\,\sum_{\alpha = 1}^{M}r_{\alpha}B_{\alpha}
  -{1\ov t}\,\mbox{diag}\,\left(\sum_{\alpha, \gamma = 0}^{M}
  B_{\alpha}B_{\gamma}\sigma_{3} -{n\ov 2}
  \sum_{\alpha = 0}^{M}B_{\alpha}(\sigma_{3}+ I)\right)
   + \Omega_{2}.\label{Gammadiag2}
  \eae
  We also  made use of the  identities,
  \[
  \mbox{diag}\,(PL) = 0 \quad \mbox{if} \quad
  \mbox{diag}\,P = 0 \quad \mbox{and} \quad \mbox{diag}\,L = L,
  \]
  and
  \be \label{diagB}
  \mbox{diag}\,\left(\sum_{\alpha = 0}^{M}
  B_{\alpha} -{n\ov 2}\,\sigma_{3}
  - {n\ov 2}\,I\right) = 0,
  \ee
  (The latter  follows from (\ref{trace1}).)

  We are at last ready to evaluate $\partial{\log D_{n}}/\partial{t}$ in
  terms of $B_{\alpha}$. To this end it is convenient
  to use
  \[
  \mbox{trace}\, \Gamma_{1} =0 \quad (\mbox{which follows from}\>
  \det Z\equiv 1)
  \]
  to rewrite  (\ref{tdk2}) in the form
  \be \label{tdk7}
  {\partial \ov \partial{t}}\log D_{n}(t) = -{1\ov 2}\mbox{trace}\,
  \left(\Gamma_{1}
  \sigma_{3}\right),
  \ee
   and then use (\ref{Gammadiag2}) and (\ref{diagB})  to obtain
  \be \label{tau}
  {\partial \ov \partial{t}}\log D_{n}(t) =
  {1\ov 2} \sum_{\alpha =0}^{M} r_{\alpha}\mbox{trace}(
   B_{\alpha}\sigma_{3}) + {1\ov 2t}\, \sum_{\alpha, \gamma = 0}^{M}
  \mbox{trace}B_{\alpha}B_{\gamma} - {n^{2}\ov 2 t}
  -{1\ov 2}\sum_{\alpha =1}^{M}r_{\alpha}k_{\alpha}.
  \ee
  For future comparison with the Jimbo-Miwa-Ueno $\,\tau$-function
  it is convenient to use (\ref{diagB}) one more time and rewrite
  (\ref{tau}) as
  \be \label{tau1}
  {\partial \ov \partial{t}}\log D_{n}(t) =
  {1\ov 2} \sum_{\alpha =0}^{M} r_{\alpha}\mbox{trace}(
   B_{\alpha}\sigma_{3}) +{1\ov 2t}\, \sum _{j\neq i = 1,2}
   (\sum_{\alpha = 0}^{M}B_{\alpha})_{ij}
  (\sum_{\gamma = 0}^{M}B_{\gamma})_{ji}
  -{1\ov 2}\sum_{\alpha =1}^{M}r_{\alpha}k_{\alpha}.
  \ee

  Let us now perform the similar transformations with the
  right hand side of equation (\ref{rdk6}). We first notice
  that its subscripts - free form can be written down as
  \be \label{rdk7}
  {\partial \ov \partial{r_{\alpha}}}\log D_{n}(t)
  =  \mbox{trace}\,
  \left(Z^{-1}(r_{\alpha})Z'(r_{\alpha})E_{\alpha}\right),
  \ee
  where
  \[
  E_{\alpha} = \left(\begin{array}{cc}
  0 & 0\\
  0 & -k_{\alpha}
  \end{array}\right)
  \]
  is the formal monodromy exponent at $r_{\alpha}$ (see (\ref{fmon}))
  and in transforming (\ref{rdk6}) into (\ref{rdk7}) we
  took into account that $\det Z(z) \equiv 1$. Secondly, by rewriting
  equation (\ref{B}) as the equation
  \[
  Z^{-1}(z)Z'(z)
  = Z^{-1}(z)B(z)Z(z) - \Omega(z),
  \]
  we get the following representation of the product
  $Z^{-1}(r_{\alpha})Z'(r_{\alpha})E_{\alpha}$,
  \bae
   Z^{-1}(r_{\alpha})Z'(r_{\alpha})E_{\alpha}
   &=& {t\ov 2}Z^{-1}(r_{\alpha})\sigma_{3}Z(r_{\alpha})E_{\alpha}
   + \sum_{{\gamma=0\atop \gamma\neq\alpha}}^{M}
   {{Z^{-1}(r_{\alpha})B_{\gamma}Z(r_{\alpha})E_{\alpha}}
   \ov{r_{\alpha}-r_{\gamma}}} \nonumber\\
   &-& {t\ov 2}\sigma_{3}E_{\alpha} -
   \sum_{{\gamma=0\atop \gamma\neq\alpha}}^{M}
   {{E_{\gamma}E_{\alpha}}
   \ov{r_{\alpha}-r_{\gamma}}}
   +[Z^{-1}(r_{\alpha})Z'(r_{\alpha})E_{\alpha}, E_{\alpha}].\label{rdk8}
  \eae
  (For notational convenience we set, as before, $r_{0}:=0$ and
  $k_{0} := -n -k$.)

  Using equation (\ref{rdk8}) in the right hand side of
  equation (\ref{rdk7}) and taking into account that
  \[
  B_{\alpha} = Z(r_{\alpha})E_{\alpha}Z^{-1}(r_{\alpha}),
  \]
  we arrive to the following $r_{\alpha}$-analog of (\ref{tau})
  \bae
  {\partial \ov \partial{r_{\alpha}}}\log D_{n}(t)&=&
   {t\ov 2}\mbox{trace}(B_{\alpha}\sigma_{3})
  + \sum_{{\gamma=0\atop \gamma\neq\alpha}}^{M}
   {{\mbox{trace}\left(B_{\alpha}B_{\gamma}\right)}
   \ov{r_{\alpha}-r_{\gamma}}} \nonumber\\
  &&\; -{k_{\alpha}t\ov 2}
  - \sum_{{\gamma=0\atop \gamma\neq\alpha}}^{M}
   {{k_{\alpha}k_{\gamma}}
   \ov{r_{\alpha}-r_{\gamma}}}, \label{tau2}
  \eae
  Combining equations (\ref{tau1}) and (\ref{tau2}) we obtain
  the main result of this section which is the following
  equation for the total differential of the function
  $\log D_{n}(t, r_{1}, ..., r_{M})$,
  \bae
  d\log D_{n} &=&  {1\ov 2}\sum_{{\alpha, \gamma=0\atop \alpha\neq\gamma}}^{M}
   \mbox{trace}\left(B_{\alpha}B_{\gamma}\right)
   {{dr_{\alpha} - dr_{\gamma}}\ov{r_{\alpha}-r_{\gamma}}}
  +{1\ov 2} \sum_{\alpha =0}^{M} \mbox{trace}(
   B_{\alpha}\sigma_{3})d(r_{\alpha}t) \nonumber \\
   &&\; +{1\ov 2}\sum _{{1\le i,j\le 2\atop i\neq j}}
   (\sum_{\alpha = 0}^{M}B_{\alpha})_{ij}
  (\sum_{\gamma = 0}^{M}B_{\gamma})_{ji} {dt\ov t} \nonumber\\
  &&\; - {1\ov 2}\sum_{\alpha =1}^{M}k_{\alpha}d(r_{\alpha}t)
  - {1\ov 2}\sum_{{\alpha, \gamma=0\atop \alpha\neq\gamma}}^{M}
   k_{\alpha}k_{\gamma}{\,{dr_{\alpha} - dr_{\gamma}}
   \ov{r_{\alpha}-r_{\gamma}}}. \label{taufinal}
   \eae
  Equation (\ref{taufinal}) describes the
  Toeplitz determinant $D_{n}(t)$ in terms of the solution
  of the Schlesinger system
  ((\ref{schles4}), (\ref{schles5a})--(\ref{schles5c}))
  up to a multiplicative constant (depending on $n$ and $k_{\alpha}$).
  Simulteneously, this equation shows, upon comparison with
  the expression (5.17) in \cite{jmu} for the logarithmic derivative
  of the $\tau$-function, that
  \be \label{tau3}
  D_{n}(t)\, =\, e^{-{t\ov 2}\sum_{\alpha}r_{\alpha}k_{\alpha}}
  \prod_{{\alpha, \gamma=0\atop \alpha\neq\gamma}}^M|r_{\alpha} - 
  r_{\gamma}|^{-{{k_{\alpha}k_{\gamma}}\ov 2}}
  \, \,\tau_{\textrm{\tiny{JMU}}},
  \ee
  where we use the notation
  $\tau_{\textrm{\tiny{JMU}}} \equiv \tau_{\textrm{\tiny{JMU}}}(t, r_0, r_{1}, ..., r_{M})$
  for the Jimbo-Miwa-Ueno
  $\,\tau$-function  corresponding to
  the linear system (\ref{fucsh})
   and evaluated for the monodromy data
  given in (\ref{fmon}) and (\ref{connect}).

\textit{Remark}. It follows from (\ref{tau3}) that $\tau_{\textrm{\tiny{JMU}}}$
vanishes as $r_\al\ra r_\gamma$ for some pair $(\al,\gamma)$.
This fact, of course, can be established directly from the definition
of the $\,\tau$-function.

  \section{Summary of the results}
  \setcounter{equation}{0}
   Recall that
  $D_{n}(\phi)$
  denotes
  the Toeplitz determinant associated with the
  symbol
  \[
  \phi(z) = e^{tz}\prod_{\alpha = 1}^{M}\left({z-r_{\alpha}
  \ov{z}}\right)^{k_{\alpha}},
  \quad -1 < r_{\alpha} < 0, \quad \alpha = 1,..., M,
  \quad r_{\alpha} \neq r_{\beta},
  \, \, \alpha \neq \beta ,
  \]
  \[
  \quad k_{\alpha} \in \textbf{N},\quad \sum k_{\alpha} = k,
  \quad t \in \textbf{R},
  \]
  and that the generating function $G_{I}(n;\{p_{i}\}, t)$
  is given by the formula
  \[
  G_{I}(n;\{p_{i}\}, t) = D_{n}(\phi), \quad r_{\alpha} = - p_{i_{\alpha}}.
  \]
  We also denote by $D_{n+1}(\phi |z)$ the Toeplitz determinant $D_{n+1}(\phi)$
  whose last row is replaced by the row $(1, z, z^{2}, ..., z^{n})$, and
  we shall assume that
  $ D_{n}(\phi) \neq 0$.

  The following theorem identifies
  $D_{n}(\phi)$ as an object of the theory of integrable
  systems; more specifically, as an object of the theory
  of generalized Schlesinger equations developed in
  \cite{jmu,jimbo}.

  \textbf{Theorem 1}. {\it Let $Z$ denote the $2\times 2$ matrix function
  defined by
  \be \label{Zf}
  Z(z)=\left(\begin{array}{cc}
  {{D_{n+1}(\phi |z)}\ov {D_{n}(\phi)}} &
   - {i \over{2\pi}}\int_{C}{{D_{n+1}(\phi |z')}\ov {D_{n}(\phi)}}
   (z')^{-n}\phi(z'){dz'\over{z-z'}} \\
  {} & {} \\
  -{{{\bar{D}}_{n}({\bar{\phi}} |1/{\bar{z}})}
  \ov {{\bar{D}}_{n}({\bar{\phi}})}}z^{n-1} &
  {i \over{2\pi}}\int_{C}{{{\bar{D}}_{n}({\bar{\phi}} |1/{\bar{z'}})}
  \ov {{\bar{D}}_{n}({\bar{\phi}})}}
  (z')^{-1}\phi(z'){dz'\over{z-z'}}\end{array}\right),
  \ee
  where $C$ is the unit circle $|z| = 1$ oriented counterclockwise.
  Introduce the $2\times 2$ matrices $B_{\alpha} :=
  B_{\alpha}(t) := B_{\alpha}(\{r_{\alpha}\},t),\quad
  \alpha =0, 1, ..., M$, by the equations,
  \be \label{Baf}
  B_{\alpha} =
  -Z(r_{\alpha})\left(\begin{array}{cc}
  0 & 0 \\
  0 & k_{\alpha} \end{array}\right)Z^{-1}(r_{\alpha}),\quad \alpha = 0, ..., M,
  \ee
  where
  \[
  r_{0} := 0, \quad \mbox{and} \quad k_{0} := -n-k.
  \]
  (The invertibility of $Z$ follows from statement 4 below.)
  Then the following statements hold:
  \begin{enumerate}
  \item
  \bae
  d\log D_{n}(t, r_{1}, ..., r_{M}) &=& 
  {1\ov 2}\sum_{{\alpha, \gamma=0\atop \alpha\neq\gamma}}^{M}
   \mbox{trace}\left(B_{\alpha}B_{\gamma}\right)
   {{dr_{\alpha} - dr_{\gamma}}\ov{r_{\alpha}-r_{\gamma}}}
  +{1\ov 2} \sum_{\alpha =0}^{M} \mbox{trace}(
   B_{\alpha}\sigma_{3})d(r_{\alpha}t) \nonumber \\
   &&\; +{1\ov 2}\sum _{{1\le i,j\le 2\atop i\neq j}}
   (\sum_{\alpha = 0}^{M}B_{\alpha})_{ij}
  (\sum_{\gamma = 0}^{M}B_{\gamma})_{ji} {dt\ov t} \nonumber\\
  &&\; - {1\ov 2}\sum_{\alpha =1}^{M}k_{\alpha}d(r_{\alpha}t)
  - {1\ov 2}\sum_{{\alpha, \gamma=0\atop \alpha\neq\gamma}}^{M}
   k_{\alpha}k_{\gamma}{{dr_{\alpha} - dr_{\gamma}}
   \ov{r_{\alpha}-r_{\gamma}}}. \label{tauf}
   \eae
   \item The matrices $B_{\alpha}$ satisfy the system of  nonlinear PDEs
  (generalized Schlesinger equations),
  \bae
  {\pl B_{\alpha} \ov {\pl t}} &=& {{n-t\, r_{\alpha}}\ov 2t}\,
  [ B_{\alpha},\sigma_{3}]
   + \sum_{\gamma =0}^{M}{[B_{\gamma}, B_{\alpha}]\ov t},\quad
  \alpha = 0, 1,\ldots, M.\label{schlesf1}\\
  {{{\pl}B_{\alpha}}\ov {{\pl}r_{\gamma}}}
  &=& {[B_{\alpha}, B_{\gamma}]\ov{r_{\alpha} - r_{\gamma}}},
  \quad \alpha \neq \gamma = 1,\ldots, M,\label{schlesf2}\\
  {{{\pl}B_{0}}\ov {{\pl}r_{\alpha}}}
  &=& {[B_{0}, B_{\alpha}]\ov{r_0 - r_{\alpha}}},\\
  {{{\pl}B_{\alpha}}\ov {{\pl}r_{\alpha}}}
  &=& \sum_{\gamma \neq \alpha}{[B_{\gamma}, B_{\alpha}]\ov
  {r_{\gamma} - r_{\alpha}}},
  \quad \alpha = 1,\ldots, M.\label{schlesf3}
  \eae
  \item Equations (\ref{schlesf1})--(\ref{schlesf3})
  are the compatibility conditions for the
  system of linear equations,
  \bae
  {{\pl \Phi(z)}\ov {\pl z}} &=&\left( {t \ov 2}\,\sigma_{3}  +
  \sum_{\alpha = 0}^{M}
  {B_{\alpha}\ov {z-r_{\alpha}}}\right)\Phi(z),\label{laxpairz}\\
  {{\pl \Phi(z)}\ov {\pl t}} &=&\left({z \ov 2}\sigma_{3} -
  {n\ov 2t}\,\sigma_{3} - {n\ov 2t}\,I
  + {1\ov t}\, \sum_{\alpha=0}^{M} B_{\alpha}\right)\Phi(z),\label{laxpairt}\\
  {{\pl \Phi(z)}\ov {\pl r_{\alpha}}}
  &=& -{{B_{\alpha}}\ov{z-r_{\alpha}}}\Phi(z),\quad
  \alpha = 1,\ldots, M,\label{laxpairr}
  \eae
  which in turn implies that the system (\ref{schlesf1})--(\ref{schlesf3})
  describes the isomonodromy deformations of the $z$-- equation
  (\ref{laxpairz}).
  \item The function $Z$ is  alternatively defined as
  an unique solution of the matrix Riemann-Hilbert problem,
  \begin{itemize}
  \item $Z$ is holomorphic for all $z\notin C$,
  \item $Z(z)z^{-n\sigma_{3}} \rightarrow I,\quad z\rightarrow \iy$,
  \item $ Z_{+}(z) = Z_{-}(z)\left( \begin{array}{cc}
  1 & z^{-n}\phi(z)\\
  0 & 1
  \end{array}\right), \quad z \in C$.
  \end{itemize}
  (In particular, we have that $\det Z\equiv 1$.)
  Equation (\ref{tauf}) can be rewritten in terms of $Z$
  as
  \be \label{tauf2}
  d\log D_{n} = -\Big(\mbox{res}_{z=\infty}(z^{n}Z_{22}(z))\Big)dt
  - \sum_{\alpha = 1}^{M}k_{\alpha}\Big (Z_{11}(r_{\alpha})Z'_{22}(r_{\alpha})
- Z_{21}(r_{\alpha})Z'_{12}(r_{\alpha})\Big )dr_{\alpha}.
  \ee
  Also,
  \be \label{tauf3}
  {D_{n+1}\ov D_{n}} = Z_{12}(0).
  \ee
  \item The function
  \[
  \Phi(z) := Z(z) e^{{tz \ov{2}}\sigma_{3}}
  \left(\begin{array}{cc}
  1 & 0 \\
  0 & z^{n}{\psi}^{-1}(z)
  \end{array}\right), \quad
  \psi(z) = \prod_{\alpha = 1}^{M}\left({z-r_{\alpha}
  \ov{z}}\right)^{k_{\alpha}},
  \]
  satisfies the linear system (\ref{laxpairz})--(\ref{laxpairr})
  with the matrices $B_{\alpha}$ given by
  (\ref{Baf}).
  \item The matrices $B_{\alpha}$ are alternatively defined
  as the solution of the inverse monodromy problem for the
  linear equation (\ref{laxpairz}) characterized by the following
  monodromy data:
  \begin{itemize}
  \item the formal monodromy exponents at the singular
  points $r_{\alpha}$, $\infty$ are given by the equations
  $$
  E_{\alpha} =
  \left(\begin{array}{cc}
  0 & 0 \\
  0 & -k_{\alpha}
  \end{array}\right),\quad
  E_{\infty} =
  \left(\begin{array}{cc}
  \hspace{-1.4ex}-n & 0 \\
  0 & 0
  \end{array}\right),
  $$
  \item the corresponding connection matrices are
  $$
  C_{\alpha} = \left(\begin{array}{cc}
  1 & \hspace{-1.2ex}-1 \\
  0 & 1
  \end{array}\right),\quad
  C_{\infty} = I.
  $$
  \item the Stokes matrices at the irregular singular point, $z=\infty$,
  are trivial.
  \end{itemize}
  \item
  \be \label{tauD}
  D_{n}(\phi)\, =\, e^{-{t\ov 2}\sum_{\alpha}r_{\alpha}k_{\alpha}}
  \prod_{{\alpha, \gamma=0\atop \alpha\neq\gamma}}^M
  |r_{\alpha} - r_{\gamma}|^{-{{k_{\alpha}k_{\gamma}}\ov 2}}
  \, \,\tau_{\textrm{\tiny{JMU}}},
  \ee
  where $\tau_{\textrm{\tiny{JMU}}} \equiv \tau_{\textrm{\tiny{JMU}}}(t, r_{0},r_{1}, ..., r_{M})$ denotes
  the Jimbo-Miwa-Ueno $\,\tau$-function  corresponding to
  the linear system (\ref{laxpairz}) and evaluated for the monodromy data
  indicated. Equation (\ref{tauD}) in turn implies the following
  representation for the generating function $G_{I}(n;\{p_{i}\}, t)$,
  \be \label{tauG}
  G_{I}(n;\{p_{i}\}, t)\, =\, e^{{t\ov 2}}
  \prod_{{\alpha, \gamma=0\atop \alpha\neq\gamma}}^M
  |p_{i_{\alpha}} - p_{i_{\gamma}}|^{-{
  {k_{\alpha}k_{\gamma}}\ov 2}}c
  \, \,\tau_{\textrm{\tiny{JMU}}}(t, 0,-p_{i_{1}},..., -p_{i_{M}}), \quad p_{i_{0}}:=0.
  \ee
  \end{enumerate} }

  {\it Remark 1\/}. In the uniform case, i.e. when $M =1$ and $k_{1} =k$,
  the linear system (\ref{laxpairz}) reduces to the $2\times 2$
  system of linear ODEs which has two regular singular
  points and one irregular point of Poincar\'e index 1. In this
  case, as it is shown in \cite{jmu, jimbo},
  the isomonodromy equations (\ref{schlesf1})--(\ref{schlesf3})
  reduce to the special case of the fifth Painlev\'e equation.
  Consequently this suggests that the uniform generating function
  $G_{I}(n; t)$ can be expressed in terms of a solution of the fifth Painlev\'e
  equation. That this is so  was obtained earlier in \cite{tw2} via
  a direct analysis of the Toeplitz determinant $D_{n}(t)$.

  {\it Remark 2\/}. The methods developed in this paper can be
  easily generalized to any symbol $\phi(z) :=
  \phi(z,t)$ such that
  \[
  \pl_{z}\log \phi,\quad \pl_{t}\log \phi
  \]
  are rational in $z$.

  {\it Remark 3\/}. In virtue of the Fredholm determinant formula
  (\ref{tfred}) for the Toeplitz determinant $D_{n}(\phi)$, equation
  (\ref{tauD}) can be interpreted as an example of the general relation
  \cite{p}
  between the Jimbo-Miwa-Ueno isomonodromy $\,\tau$-function and
  the Sato-Segal-Wilson $\,\tau$-function defined via an appropriate
  determinant bundle
  (see also \cite{hi} for another example of this  relation).

  {\it Remark 4\/}. The generalized Schlesinger system
  (\ref{schlesf1})--(\ref{schlesf3}) appeared earlier in \cite{jmms}
  in connection with the sine kernel Fredholm determinant considered
  on a union of intervals.
  The corresponding monodromy data, and hence the solution,
  are different from the ones related to the Toeplitz
  determinant $D_{n}(t)$. For instance, the sine kernel
  monodromy matrices are not trivial (see \cite{jmms};
  see also \cite{hi} for higher matrix dimensional
  generalizations); in fact, each of them equals
  the identity matrix plus a one dimensional projection.

  {\it Remark 5\/}. From the point of view of
  the asymptotic analysis of the Toeplitz determinant, the
  most important statement of Theorem 1 is Statement 4. It
  allows one to apply the Riemann-Hilbert asymptotic
  methods of \cite{dz,bi, bdj1}.

\textit{Remark 6}. This paper has been primarily concerned with
the isomondromy/Riemann-Hilbert aspect of our
 integrable system.  Presumably an analysis of the additional
 compatibility conditions,
 which arise if one extends (\ref{laxpairz})--(\ref{laxpairr})
 by a relevant $n$-difference equation, 
  would lead to a Toda like system, 
 see Okounkov~\cite{okounkov} and Adler and van Moerbeke~\cite{adler1, vanM}.
 
 \vspace{3ex}
\textbf{\large Acknowledgments} 
  
This work was begun during the MSRI Semester
Random Matrix Models and Their Applications.  We wish to thank
D.~Eisenbud and H.~Rossi for their support during this semester.
This work was supported in part by the National Science Foundation
through grants DMS--9801608, DMS--9802122 and DMS--9732687.  
The last two authors thank Y.~Chen for his kind
hospitality at Imperial College where part
of this work was done as well as the  EPSRC for the award
of a Visiting Fellowship, GR/M16580, that made this visit possible. 
We are also grateful to the referee for several valuable suggestions.


\begin{thebibliography}{99}
  
  \bibitem{adler1} M.~Adler and P.~van Moerbeke,
  Integrals over classical groups, random permutations,
  Toda and Toeplitz lattices, preprint (arXiv: math.CO/9912143).
  
 

  \bibitem{aldous} D.~Aldous and P.~Diaconis, Hammersley's interacting
  particle process and longest increasing subsequences,
  \textit{Probab.\ Theory Related Fields} \textbf{103} (1995),
  199--213.

  \bibitem{bdj1} J.~Baik, P.~Deift, and K.~Johansson,
  On the distribution of the length of the longest increasing
  subsequence of random permutations, \textit{J.\  Amer.\  Math.\
  Soc.} \textbf{12} (1999), 1119--1178.

  \bibitem{bw} E.~L.~Basor and H.~Widom, On a Toeplitz determinant identity
  of Borodin and Okounkov, preprint (arXiv: math.FA/9909010).

   \bibitem{bayer} D.~Bayer and P.~Diaconis, Trailing the dovetail
   shuffle to its lair, \textit{Ann.\  Applied Probability} \textbf{2}
   (1992), 294--313.

  \bibitem{bi} P.~M.~Bleher and A.~R.~Its,
  Semiclassical asymptotics of orthogonal polynomials, Riemann-Hilbert
  problem, and universality in the matrix model,
  {\it  Ann.\ of Math.\/} {\bf 150} (1999), 185--266.


  \bibitem {borodin} A.~Borodin, Rimann-Hilbert problem and the discrete
  Bessel kernel, preprint (arXiv: math.CA/9912093).

  \bibitem{bo} A.~Borodin and A.~Okounkov, A Fredholm determinant formula
  for Toeplitz determinants, preprint (arXiv: math.CA/9907165).


  \bibitem{boo} A.~Borodin, A.~Okounkov and G.~Olshanski,
  Asymptotics of Plancherel measures for symmetric groups,
  \textit{J.\ Amer.\ Math.\ Soc.} \textbf{13} (2000), 481--515.

  \bibitem{deift} P.~A. Deift, Integrable operators, in
  \textit{Differential operators and spectral theory: M. Sh. Birman's
  70th anniversary collection}, V.~Buslaev, M.~Solomyak, D.~Yafaev, eds.,
  American mathematical Society Translations, ser.~2, v.~189, Providence,
  RI: AMS, 1999.


  \bibitem{dz} P.~A.~Deift and X.~Zhou, A steepest
  descent method for oscillatory Riemann-Hilbert problems:
  Asymptotics for the MKdV equation, {\it  Ann.\ Math.\/} {\bf 137}
  (1993), 295--368.


  \bibitem {fik} A.~S.~Fokas, A.~R.~Its, and A.~V.~Kitaev,
  The isomonodromy approach to matrix models in 2D quantum gravity,
  {\it Commun.\  Math.\ Phys.\/} {\bf 147} (1992), 395--430.



  \bibitem{gessel} I.~M.~Gessel, Symmetric functions and
  P-recursiveness, {\it  J.\ Comb.\ Theory, Ser.~A\/} {\bf 53} (1990),
  257--285.


  \bibitem {hi} J.~Harnad and A.~R.~Its, Integrable Fredholm operators
  and dual isomonodromic deformations, preprint (arXiv: solv-int/9706002).


  \bibitem{iiks} A.~R.~Its, A.~G.~Izergin, V.~E.~Korepin, and N.~A.~Slavnov,
  Differential equations for quantum correlation functions, {\it Int.\ J.\
Mod.\
  Physics\/}  {\bf B4} (1990), 1003--1037.

  \bibitem{itw1} A.~R.~Its, C.~A.~Tracy, and H.~Widom. Random  words,
  Toeplitz determinants
  and integrable systems. I, preprint (arXiv: math.CO/9909169).

  \bibitem{jmms} M.~Jimbo, T.~Miwa, Y.~Mori, and M. Sato,
  Density matrix of an impenetrable Bose gas and the
  fifth Painlev\'e trranscendent, {\it  Physica\/} {\bf 1D} (1980), 80--158.

  \bibitem{jmu}  M.~Jimbo, T.~Miwa, and K.~Ueno,  Monodromy preserving
  deformation of linear ordinary differential equations with
  rational coefficients. I, {\it  Physica\/} {\bf 2D} (1981), 306--352.

  \bibitem{jimbo} M.~Jimbo and T.~Miwa, Monodromy preserving
  deformation of linear ordinary differential equations with
  rational coefficients. II, {\it  Physica\/} {\bf 2D} (1981), 407--448.

  \bibitem{johansson0} K.~Johansson, The longest increasing subsequence in a
random permutation and a unitary random matrix model, \textit{Math.\ Research \
Lett.}
  \textbf{5} (1998), 63--82.
  
  \bibitem{johansson1} K.~Johansson, Shape fluctuations and random
  matrices, \textit{Commun.\ Math.\ Phys.} \textbf{209} (2000), 437--476.

  \bibitem{johansson2} K.~Johansson, Discrete orthogonal polynomial
  ensembles and the Plancherel measure, preprint (arXiv: math.CO/9906120).

  \bibitem{mehta} M.~L.~Mehta, {\it Random Matrices\/},
  2nd ed.,  Academic Press, San Diego, 1991.
 
  \bibitem{nmpz} S.~Novikov and S.~V.~Manakov, L.~P.~Pitaevskii,
  V.~E.~Zakharov, \textit{Theory of Solitons. The Inverse Scattering Method},
  1984 Consultants Bureau, New York.

\bibitem{okounkov} A.~Okounkov, Infinite wedge and random partitions,
preprint (arXiv: math.RT/9907127).

  \bibitem{p} J.~Palmer, Zeros of the Jimbo, Miwa, Ueno tau function,
  \textit{J.\ Math.\ Phys.} \textbf{40} (1999), 6638--6681.

  \bibitem{stanleyBook} R.~P.~Stanley, \textit{Enumerative Combinatorics},
  Vol.~2, Cambridge University Press, 1999.

  \bibitem{stanley} R.~P.~Stanley, Generalized riffle shuffles and
  quasisymmetric functions, preprint (arXiv: math.CO/9912025).

  \bibitem{Sz1} G.~Szeg\"o, \textit{Orthogonal Polynomials}, American
Mathematical
  Society. Colloquium Publications, Vol.~ 23, 4th Ed.,  New York, 1975.

  \bibitem{tw2} C.~A.~Tracy and H.~Widom, On the distributions of the lengths
  of the longest monotone subsequences in random words, preprint (arXiv:
math.CO/9904042).

  \bibitem{tw3} C.~A.~Tracy and H.~Widom, Fredholm determinants, differential
equations
  and matrix models, {\it  Commun.\ Math.\ Phys.\/}
   {\bf 163} (1994), 33--72.


\bibitem{ueno} K.~Ueno and K.~Takasaki, Toda lattice hierarchy, in
\textit{Adv.\ Studies in Pure Math.} \textbf{4}, Group Representations
and Systems of Differential Equations, 1994, 1--95.

\bibitem{vanM} P.~van Moerbeke, Integrable lattices: random
  matrices and random permutations, preprint (arXiv: math.CO/0010135).

  \end{thebibliography}
  \end{document}